\documentstyle[epsfig]{mn}

\def\i{\item}
\def\bi{\bibitem{}}

\def\ni{\noindent}
\def\beb{\begin{thebibliography}{999}}
\def\eeb{\end{thebibliography}}
\def\bei{\begin{itemize}}
\def\eei{\end{itemize}}
\def\bef{\begin{figure}}
\def\eef{\end{figure}}
\def\ben{\begin{enumerate}}
\def\een{\end{enumerate}}
\def\beq{\begin{equation}}
\def\eeq{\end{equation}}
\def\ber{\begin{eqnarray}}
\def\eer{\end{eqnarray}}
\def\edo{\end{document}}

\def\Bb{{\bf B}}
\def\pa{\partial}
\def\vb{{\bf v}}
\def\vt{v_{\theta}}
\def\half{\frac{1}{2}}
\def\third{\frac{1}{3}}
\def\Thalf{$T_{1/2}$}

\newcommand{\gcc}{{\rm g} \, {\rm cm}^{-3}}

\newcommand{\msun}{\mbox{{\rm M}$_{\odot}$}}
\newcommand{\mdot}{\mbox{$\dot{M}$}}

\begin{document}
\title[neutron star field evolution]
{Diamagnetic Screening of the Magnetic Field in Accreting Neutron Stars}
\author[Choudhuri and Konar]
{Arnab Rai Choudhuri$^{1,{\dag}}$ and Sushan Konar$^{2,{\dag}}$ \\ 
$^1$Department of Physics, Indian Institute of Science, Bangalore 560012, India \\
$^2$Inter-University Centre for Astronomy \& Astrophysics, Pune 411007, India \\ 
$^{\dag}$ e-mail : arnab@physics.iisc.ernet.in, sushan@iucaa.ernet.in \\ }
\date{1st August, 2001}
\maketitle

\begin{abstract}
A possible mechanism for screening of the surface magnetic field of an accreting neutron star, by the accreted material,
is investigated. We model the material flow in the surface layers of the star by an assumed 2-D velocity field satisfying 
all the physical requirements. Using this model velocity we find that, in the absence of magnetic buoyancy, the surface 
field is screened (submergence of the field by advection) within the time scale of material flow of the top layers. On the 
other hand, if magnetic buoyancy is present, the screening happens over a time scale which is characteristic of the slower 
flow of the deeper (hence, denser) layers. For accreting neutron stars, this longer time scale turns out to be about $10^5$ 
years, which is of the similar order as the accretion time scale of most massive X-ray binaries.
\end{abstract}


\section{introduction}
\label{sintro}

\ni Radio pulsars can be broadly classified into two categories: (a) isolated pulsars with rotation periods usually above 
1 s and very strong magnetic fields ($10^{12}$ -- $10^{13}$~G); and (b) binary/millisecond pulsars with much shorter rotation 
periods and considerably weaker magnetic fields ($10^8$ -- $10^{10}$~G). Observational indications have led to the proposal of 
a unified model in which these two types of pulsars are regarded as two phases in the evolutionary history of neutron stars 
(see Bhattacharya 1995, 1996; van den Heuvel 1995; Verbunt \& van den Heuvel 1995 and references therein). According to this 
model, a normal pulsar, if it is in a binary system, may undergo mass accretion from its binary companion for a limited period, 
and the angular momentum of the accreted matter may spin up the neutron star, ultimately making it appear as a millisecond 
pulsar after the accretion process stops. In this scenario, neutron stars seen in the X-ray as members of binary systems should 
correspond to the intermediate phase of transition from the normal pulsar phase to the millisecond pulsar phase. The fact that 
many millisecond pulsars are found in binary systems lends credence to this picture. If this scenario is correct, then the most 
natural explanation of the weaker magnetic fields of millisecond pulsars must be that the magnetic field of a neutron star is 
significantly reduced during the accretion phase. Though, how this happens is still not understood very well. \\

\ni A variety of models have been proposed for the evolution of the magnetic field in an accreting neutron
star (see Bhattacharya 1999a, 1999b; Konar \& Bhattacharya 1999 and references therein). There are two classes 
of models which have been mainly explored in this context: one that relates the magnetic field evolution to the 
spin evolution of the star and the other attributing the field evolution to direct effects of mass accretion. 
The different classes of models usually assume different kinds of initial field configurations. Models depending 
on spin-down assume a core-flux supported by proton superconductor flux tubes. On
the other hand, models invoking ohmic 
dissipation usually assume an initial crustal configuration. The mechanism of ohmic decay, being unique to the 
crustal currents, is also used in models where spin-down is invoked for flux expulsion, for a subsequent dissipation 
of such flux in the crust (Jahan Miri \& Bhattacharya 1994; Bhattacharya \& Datta 1996, Konenkov \& Geppert 200, 2001a, 2001b). 
In an accretion-heated 
crust, the decay takes place principally as a result of rapid dissipation of currents due to the decrease in 
the electrical conductivity and hence a reduction in the ohmic dissipation time scale (Geppert \& Urpin 1994; 
Urpin \& Geppert 1995; Urpin \& Geppert 1996; Konar \& Bhattacharya 1997). \\

\ni However, in this context one of the least investigated mechanisms that has often been invoked is the possible 
screening of the field by accreted material. The idea of a possible screening of the magnetic field of a neutron star 
by accreting material was first proposed by Bisnovatyi-Kogan \& Komberg (1974). Later, Taam \& van den Heuvel 
(1980) indicated that the accreted matter, which is completely ionized plasma and hence diamagnetic in nature, 
might screen the pre-existing field, reducing its strength at the surface. Blandford, De Campli \& K\"{o}nigl 
(1979) suggested that the material accreting onto the poles will be confined by the strong magnetic stresses near 
the surface of the star. At low accretion rates, the material sinks below the stellar surface until the hydrostatic 
pressure of the stellar material is as large as the magnetic pressure. The plasma then flows sideways giving rise 
to horizontal components at the expense of the vertical field and may result in a decrease in the observed field 
strength. Then it was shown by Woosley \& Wallace (1982) and Hameury, Bonazzola, Heyvaerts \& Lasota (1983) that the 
accretion column is like a small mountain on the polar cap rather than being subsurface. Semi-quantitative calculations by 
Romani (1990, 1995) showed that hydrodynamic flows in the surface layers may bury the field to deeper layers, effectively 
reducing the surface strength. \\

\ni When the accreting material flows horizontally from the polar caps to lower latitudes, magnetic field lines are expected 
to be dragged by the flow, giving rise to distortions in the original field structure. This may lead to creation of additional
horizontal components of the field at the expense of the vertical ones. Now, horizontal magnetic fields are known to be subject 
to magnetic buoyancy (see, for example, Parker 1979, \S8.7--8). Although the flow of accreting matter tries to screen the 
magnetic field, the field may pop back to the surface due to magnetic buoyancy (Spruit 1983) and Ohmic re-diffusion. The magnetic 
loops coming out to the surface due to buoyancy would reconnect with overlying field lines and thereby would make the screening 
ineffective. Simple estimates by Konar (1997) showed that the instability or the reconnection time scales happen to be too small 
for the screening mechanism to be effective. Similar conclusion, as for the re-diffusion time-scale being of the order of 
$10^3-10^4$ years, have been reached by Geppert, Page \& Zannias (1999), in the context of the hyper-critical accretion onto a 
newly born neutron star and the re-emergence of its surface field. \\

\ni It seems at the first sight that the accreting material will not be able to screen the magnetic field of the neutron star.  
In particular, for higher rates of accretion, effects such as magnetic buoyancy and Ohmic diffusion might make the magnetic 
field emerge to the surface, rapidly, through the accreting matter.  
We, however, argue here that this is a complex problem of competing time scales and one has to do more detailed calculations,
than what has been done previously, to settle the question of 
screening of the magnetic field by accreted matter. To the 
best of our knowledge, no attempt has so far been made to investigate this problem in a 2-dimensional framework and ours is the
first 2-D calculation of the problem. Some of the results we present here depend crucially on the 2-D geometry and could not have
been obtained by simpler 1-D calculations.  \\

\ni The accreted material flowing from the polar caps would meet near the equator. This is expected to cause a submergence 
of such material under the surface over a belt around the equator. Even though it is difficult to say anything definite 
about the flow of the accreted material in the surface layers of the neutron star, without considering the Navier-Stokes
equation including a reasonable viscosity; we do not really expect a pile-up of material in the 
equatorial region. Hence we expect that the submerging material would push against the solid interior of the neutron star 
and would displace the 
solid layers. Our rough estimates suggest the speed of this displacement of the solid interior to be $\sim 10^{-6}$ cm s$^{-1}$
for Eddington accretion rates --- comparable to the speeds of tectonic plates on the Earth's surface which are again
$\sim 10^{-6}$ cm s$^{-1}$ (see, for example, Emiliani 1992, \S12.2). We propose that the solid inner material in the equatorial 
region, while remaining solid, is displaced 
very slowly downwards as well as sideways to higher latitudes in the form of a counter-flow to the equator-ward flow of 
accreted material at the surface. This very slow movement of the solid material would carry the magnetic field with it, since 
magnetic buoyancy would be ineffective in a region where the material is solid. The time taken by this very slow motions to cover 
an appreciable distance inside the neutron star is $\sim 10^5$ years. Our 2-D calculations show that, in presence of accretion,
the magnetic field can be screened in the time scale of this very slow interior motions of the solid material. Once the accretion
is turned off, the submerged field should start re-diffusing to the outer layers. \\

\ni Although our present work is the first quantitative study of the effect on magnetic field of a flow in the meridional 
plane in the context of neutron stars, the effect of a meridional flow on solar magnetic fields has been studied in some detail 
(Dikpati \& Choudhuri 1994, 1995; Choudhuri \& Dikpati 1999). Some recent models of the solar dynamo depend on the existence of
a meridional flow in the solar convection zone in a very crucial way (Choudhuri, Sch\"ussler \& Dikpati 1995; Durney 1995, 1997; 
Dikpati \& Charbonneau 1999; Nandy \& Choudhuri 2001). Our work derives its methodology from the models developed by the authors
mentioned above. \\

\ni The mathematical formulation of the problem is discussed in the next section. Although we have run our code over a wide range 
of parameters, our code cannot handle the extreme values of parameters appropriate for the crust of a neutron star. The density 
varies in the outer layers of the neutron star by 8 orders of magnitude; the accreted material flows through a layer of thickness 
equal to 1\% of the radius of the neutron star and the time scales of motion and Ohmic decay again differ by 8 orders of magnitude!  
In \S3, we present some general results obtained with more moderate values of the parameters to understand the basic physics and 
show that one can draw some fairly generic, parameter-independent conclusions. The lessons learnt from this general study are 
extrapolated to the situation of the neutron star in \S4. Finally, our conclusions are summarized in the last section.   


\section{mathematical formulation}

The magnetic field of the neutron star will evolve according to the well-known induction equation:
\beq
\frac{\pa \Bb}{\pa t} = \nabla \times (\vb \times \Bb) - \frac{c^2}{4\pi} \nabla \times (\frac{1}{\sigma} \nabla \times \Bb) \,,
\label{eq_ind}
\eeq
where $\sigma$ is the electrical conductivity of the medium (see, for example, Parker 1979; Choudhuri 1998). We assume 
an axisymmetric poloidal field which enables us to represent the magnetic field in the following form,
\beq
\Bb = \nabla \times [ A(r, \theta) \hat{{\bf e_\phi}}].
\label{eq_vec}
\eeq
It is easy to show that the magnetic field lines are given by contours of constant $r \sin \theta A$. On substituting eq.(\ref{eq_vec})
in eq.(\ref{eq_ind}), we find that $A$ evolves according to the equation:
\beq
\frac{\pa A}{\pa t} + \frac{1}{s} (\vb. \nabla)(s A) = \eta \left( \nabla^2 - \frac{1}{s^2} \right)A \,, 
\label{eq_dadt}
\eeq
where $\eta = c^2/4\pi\sigma$ and $s = r \sin \theta$. It is clear from eq.(\ref{eq_dadt}) that only the poloidal component of 
$\vb$ affects the evolution of $A$. \\

\ni To study the evolution of the magnetic field by solving eq.(\ref{eq_dadt}), we need to know the velocity field $\vb$.  
It is an extremely non-trivial problem to calculate the velocity field $\vb$ of the accreted material from the basic principles 
of fluid mechanics. We, therefore, follow the kinematic approach of specifying a velocity field $\vb$ which has the required 
characteristics and then study the effect of this velocity field on the initial magnetic field. The material accumulated in 
the polar regions would cause an equator-ward flow at the surface in both the hemispheres. Let us assume that this flow is
confined in a shell from $r = r_m$ to the surface $r = r_s$ ($r_m < r_s$), i.e, primarily in the liquid part of the crust. Near 
the equator, the flow originating from the two polar regions would meet, turn around and sink under the surface. We expect a 
pole-ward counter-flow in a region immediately below $r =r_m$. Eventually this material settles radially on the core. So we 
expect $\vb$ to become radially inward at some radius $r = r_b$ ($r_b < r_m$). The pole-ward counter-flow as well as the radially 
inward flow would take place mainly in the solid crystalline region. We have a source of new material in the polar cap region.   
Apart from that, we must have 
\beq
\nabla. (\rho \vb) = 0
\eeq
everywhere else. We now write down a velocity field which has all these characteristics. In the top layer $r_m < r < r_s$ of 
equator-ward flow, we take 
\ber
\rho v_r = K_1 \left[\frac{1}{3} r - \frac{1}{2} r_m 
             + \frac{\left( \frac{1}{2} r_m - \frac{1}{3} r_s \right) r_s^2}{r^2} \right] e^{- \beta \cos^2 \theta}  \,
\label{vr1}
\eer
\ber
\rho v_{\theta} = \half \sqrt{\frac{\pi}{\beta}} K_1 \frac{(r - r_m)}{\sin \theta} 
                    \mbox{erf} (\sqrt{\beta} \cos \theta) \left(1 - e^{-\gamma \theta^2} \right) \,. 
\label{vth1}
\eer
In the next lower layer $r_b < r <r_m$ where we expect the flow to turn around, we take
\ber
\rho v_r &=& K_2 \, e^{-\beta \cos^2 \theta} \nonumber \\
         &\times& \left[\third r - \half (r_m + r_b) + \frac{r_m r_b}{r} 
              + \frac{\left(\frac{1}{6} r_b - \half r_m \right) r_b^2}{r^2} \right] \nonumber \\
         &-& \left[\third r - \half (r_m + r_b) + \frac{r_m r_b}{r} 
             + \frac{\left( \frac{1}{6} r_m - \half r_b \right) r_m^2}{r^2} \right] \nonumber \\
         &\times& \half \sqrt{\frac{\pi}{\beta}} K_2 \,
\label{vr2}
\eer
\ber
\rho v_{\theta} &=& \frac{K_2}{2 \sin \theta} \sqrt{\frac{\pi}{\beta}} \left( r + \frac{r_m r_b}{r} - r_b - r_m \right) \nonumber \\
                && \times \left[\mbox{erf}(\sqrt{\beta} \cos \theta) - \cos \theta \right] \,.
\label{vth2}
\eer
Finally, when $r<r_b$, we take the velocity to be radially inward (characteristic of the radial compression in the deeper layers): 
\ber
\rho v_r = - \frac{K_3}{r^2} \,
\label{vr3}
\eer
\ber
\rho v_{\theta} = 0 \,.
\label{vth3}
\eer
It is evident from the context that the rate of accretion is related to these coefficients by the relation $K_3 = \mdot/{4 \pi}$.
The coefficients $K_1$, $K_2$ and $K_3$ can be related from the fact that $\rho v_r$ has to be continuous across $r = r_m$ and 
$r = r_b$. From the continuity of $\rho v_r$ at $r = r_m$, we obtain 
\ber
&& K_1 \left[ - \frac{1}{6} r_m + \frac{\left( \frac{1}{2} r_m - \frac{1}{3} r_s \right) r_s^2}{r_m^2} \right] \nonumber \\
&& = K_2 \left[ - \frac{1}{6} r_m + \half r_b + \frac{\left( \frac{1}{6} r_b - \half r_m \right) r_b^2}{r_m^2} \right],
\eer
whereas the continuity of $\rho v_r$ at $r = r_b$ gives 
\beq
\frac{K_2}{2} \sqrt{\frac{\pi}{\beta}} \left[ -\frac{1}{6} r_b + \half r_m  
+ \frac{\left( \frac{1}{6} r_m - \half r_b \right) r_m^2}{r_b^2} \right] = \frac{K_3}{r_b^2}. 
\eeq
We have specified $\rho v_{\theta}$ in such a fashion that it is equator-ward in the upper layer $r_m <r <r_s$ and pole-ward 
in the lower layer $r_b <r <r_m$, falling to zero both at $r =r_m$ and $r = r_b$.  It is straightforward to show that 
$\nabla. (\rho \vb) = 0$ everywhere below $r = r_m$. If we choose the values of $\beta$ and $\gamma$ in such a fashion that the 
factors $\mbox{erf} (\sqrt{\beta} \cos \theta)$ and $\left( 1 - e^{-\gamma \theta^2} \right)$ have values different from unity in 
the non-overlapping regions around the equator and the pole respectively, then the divergence in the uppermost layer above 
$r = r_m$ turns out to be :
\beq
\nabla. (\rho \vb) = \sqrt{\frac{\pi}{\beta}} \gamma K_1 \left(1 - \frac{r}{r_m} \right)
                     \frac{\theta}{\sin \theta} e^{- \gamma \theta^2}.
\eeq
We thus have a source of material only in the region around the pole in the upper layer. Our expression for the velocity field 
given by (\ref{vr1})--(\ref{vth3}) may seem somewhat complicated to the reader. But this is the simplest velocity field 
satisfying all the necessary characteristics that we have been able to write down. \\

\bef
\begin{center}{\mbox{\epsfig{file=fig01.ps,width=200pt,angle=-90}}}\end{center}
\begin{center}{\mbox{\epsfig{file=fig01a.ps,width=200pt,angle=-90}}}\end{center}
\caption[]{The top panel shows $\rho \vb$ in a right-angular slice ($0 \leq \theta \leq \pi/2$, 
$0.25 \leq r \leq 1.0$). We have used $r_m = 0.75$, $r_b = 0.5$ in this picture. The bottom panel shows the contours of
$\nabla \cdot({\rho \vb})$ for this flow velocity. Note that the divergence vanishes everywhere except in a narrow
region near the polar cap.}
\label{fig1}
\eef

\ni Fig.~\ref{fig1} shows how $\rho \vb$, given by eqs.(\ref{vr1})--(\ref{vth3}), looks like. In all our discussions, we shall 
assume the radius $r_s$ of the neutron star to be the unit of length. Fig.~\ref{fig1} has been generated by taking $r_m = 0.75$, 
$r_b = 0.5$, $\beta = 1.0$ and $\gamma = 1.0$. If we assume the density to be constant, then the amplitude of the equator-ward 
flow in the top layer and the amplitude of the pole-ward counter-flow in the layer underneath it are of comparable magnitude.
But in the crustal layers of a neutron star, the density varies by 8 orders of magnitude, which is difficult for a numerical code 
to handle. Therefore, to study the effect of density stratification, we allow the density to vary by 2 orders. This is done by 
modeling the density variation with the radius as
\beq
\rho (r) = 1.01 - e^{-\alpha (r_s - r)},
\label{eq_rho}
\eeq
with $\alpha = 5.0$. A plot of this variation of density with radius is shown in Fig.~\ref{fig2}. This figure also shows the 
variation of $\vt$ as a function of $r$ at the mid-latitude ($\theta = \pi/4$). The unit of velocity is set by equating $K_3$ 
appearing in eq.(\ref{vr3}) to unity. This, along with the unit of length defined by the radius of the neutron star, fixes the 
unit of time in our problem. So, the unit of time is essentially the convective time scale from eq.(\ref{eq_dadt}), given by
\beq
\tau_{\rm conv} \sim \frac{v}{r} \,.
\eeq
We see in Fig.~\ref{fig2} that the equator-ward flow velocity near the surface is 
about 100.0. The time scale for the flow (i.e.\ the time taken by this flow to traverse the radius of the neutron star) is 
about 0.01 unit of time. The pole-ward counter-flow under the top layer has an amplitude of about $-2.0$, giving a time scale of
 about 0.5. This flow is weaker by a factor of about 100 compared to the flow near the surface because the density in this deeper 
layer is enhanced by a factor of 100 with respect to the density at the surface. \\

\bef
\begin{center}{\mbox{\epsfig{file=fig02.ps,width=300pt,angle=-90}}}\end{center}
\caption[]{Radial dependence of density and horizontal velocity $v_\theta$ at mid-latitude. 
As we move deeper below the surface, note that $v_\theta$ becomes smaller, changes sign and then drops to
zero as the flow velocity becomes increasingly radial in the interior regions.}
\label{fig2}
\eef

\ni To obtain the evolution of the magnetic field with time, we need to solve eq.(\ref{eq_dadt}) with the velocity field defined
by eq.(\ref{vr1})-(\ref{vth3}). We integrate eq.(\ref{eq_dadt}) numerically in the region $0<\theta<\pi/2$ and $r_0 <r <r_s$, 
where the interior boundary of integration $r_0$ is taken some distance below $r_b$. 
The following boundary conditions are satisfied.
The field lines from the two hemispheres should match smoothly at the equator, which
requires 
\beq
\frac{\partial A}{\partial \theta} = 0 \,, \, \mbox{at} \, \, \theta = \pi/2. 
\eeq
To avoid a singularity at the pole, we need to have
\beq
A = 0 \,, \, \mbox{at} \, \, \theta = 0.
\eeq
Now, at the upper boundary $r = r_s$, we allow the magnetic field to smoothly match a potential field outside (Dikpati \& 
Choudhuri 1994), which satisfies the free space Laplace equation 
\beq
\left(\nabla^2 - \frac{1}{r^2 \sin^2 \theta} \right) A = 0 \,.
\eeq
Its solution has the general form
\beq
A(r \ge r_s, \theta, t) = \sum_n \frac{a_n(t)}{r^{n+1}} P^l_n(\cos \theta),
\eeq
which, to match the boundary condition at $\theta = \pi/2$ mentioned above, picks up only the odd $n$ terms.
The coefficients $a_n$ are obtained from the surface values of $A$ by the following relation
\beq
a_n(t) = \frac{2n+1}{n(n+1)} {r_s}^{n+1} \int^{\pi/2}_0 A(r_s,\theta,t) P^l_n(\cos \theta) \sin \theta d\theta \,.
\eeq
The derivative of $A$ at the surface is
\beq
\frac{\partial A}{\partial r}\Big|_{r=r_s} = - \sum_n (n+1) \frac{a_n(t)}{{r_s}^{n+2}} P^l_n(\cos \theta) \,.
\eeq
Hence, the top boundary condition is that our solution has to match (21) at $r = r_s$,
with $a_n(t)$ being given by (20).  Finally,
at the lower boundary $r = r_0$, we allow the magnetic field to be carried freely downward by the radially inward velocity 
field there. How this is implemented in the numerical scheme is
pointed out in the Appendix.
%
%
%
%

Our code is developed by using the code of Dikpati \& Choudhuri (1994) developed in the context of solar MHD as a template. 
We summarize the salient features of the numerical schemes in the Appendix of the present paper.  

\section{field evolution - general results}

\ni So far, the evolution of a magnetic field due to a flow like what we are considering has 
not been studied before. Therefore, to begin with, we present a general study of the problem, to gain an understanding of it, 
before we consider applications to the neutron star. In a neutron star, the velocity field remains confined to the uppermost 
layer, which is rather thin compared to the size of the star (the equator-ward flow takes place in a layer of 100 m, which is 
about 1\% of the neutron star radius). Solving the equations numerically in such a thin layer involves dealing with special 
numerical instabilities. Nor is it easy to plot the resulting magnetic field configurations without making some transformations 
to blow up the thin region where all action takes place.  We, therefore, first study the basic physics of the problem by taking 
the velocity field confined to a reasonably thick layer. Most of the calculations presented in this section uses the velocity 
field shown in Fig.~\ref{fig1}. We shall see that it is possible to draw important conclusions which are independent of the 
thickness of the flow region. The implications of our calculations for an accreting neutron 
star will be discussed in the next section.

\subsection{A Central Dipole}

\bef
\begin{center}{\mbox{\epsfig{file=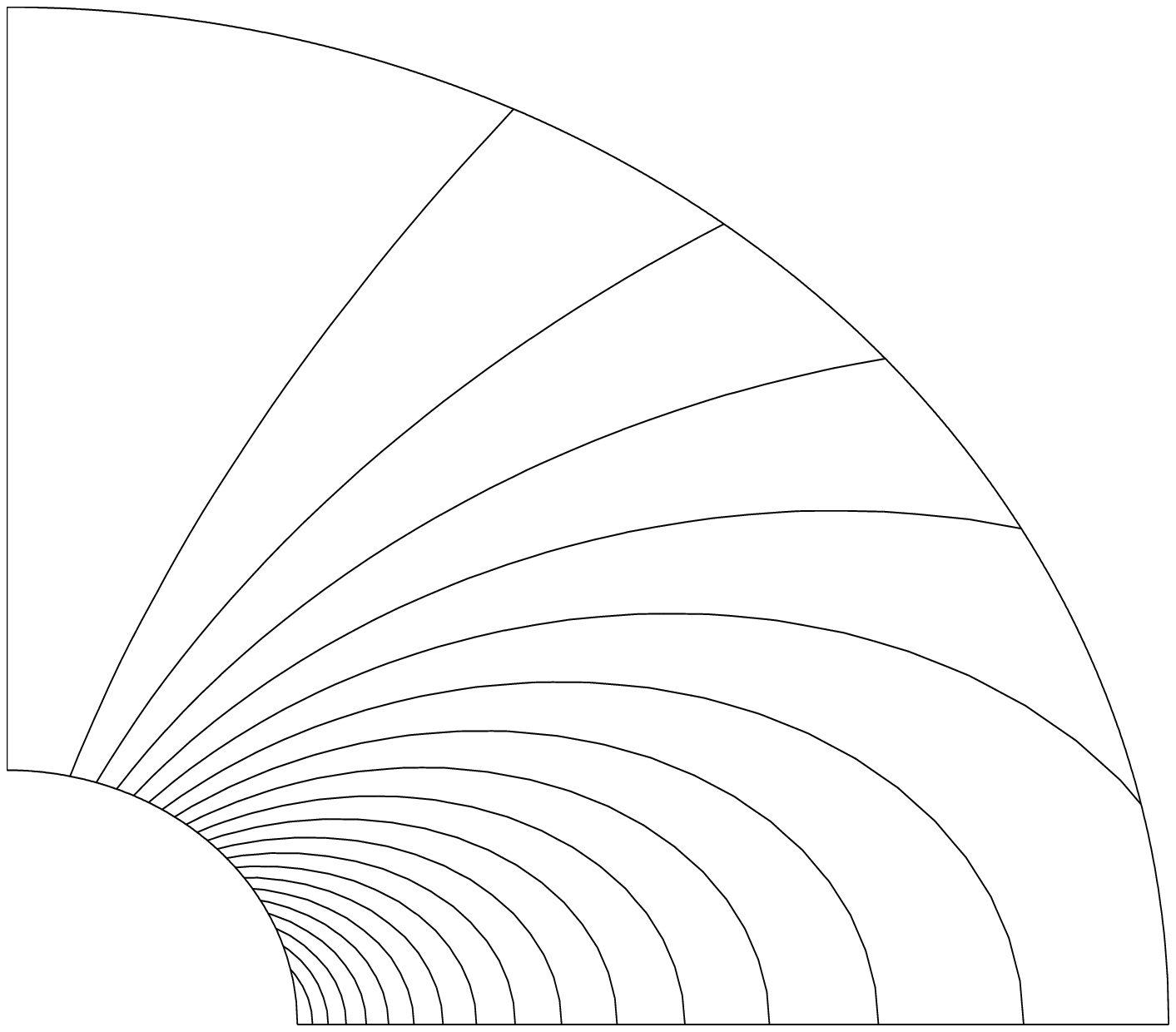,width=250pt}}}\end{center}
\caption[]{Initial field configuration, assuming a central dipole (given by eq.(\ref{eq_dip})), in a region defined by 
$0.25 \leq r \leq 1$ and $0 \leq \theta \leq \pi/2$.}
\label{fig3}
\eef

\ni There is no clear consensus regarding the question as to whether the magnetic field of a typical neutron star penetrates 
through the core or whether it is confined to the crust. We shall first present results taking the initial 
magnetic field to be in the form of a dipolar
potential field centred at the centre of the star and pervading the entire star. Fig.~\ref{fig3} shows the field lines for 
such a potential field, given by
\beq
A = \frac{\sin \theta}{r^2};
\label{eq_dip}
\eeq
which we assume to be the initial configuration of the field. 

\subsubsection{Without Magnetic Buoyancy}

\bef
\begin{center}{\mbox{\epsfig{file=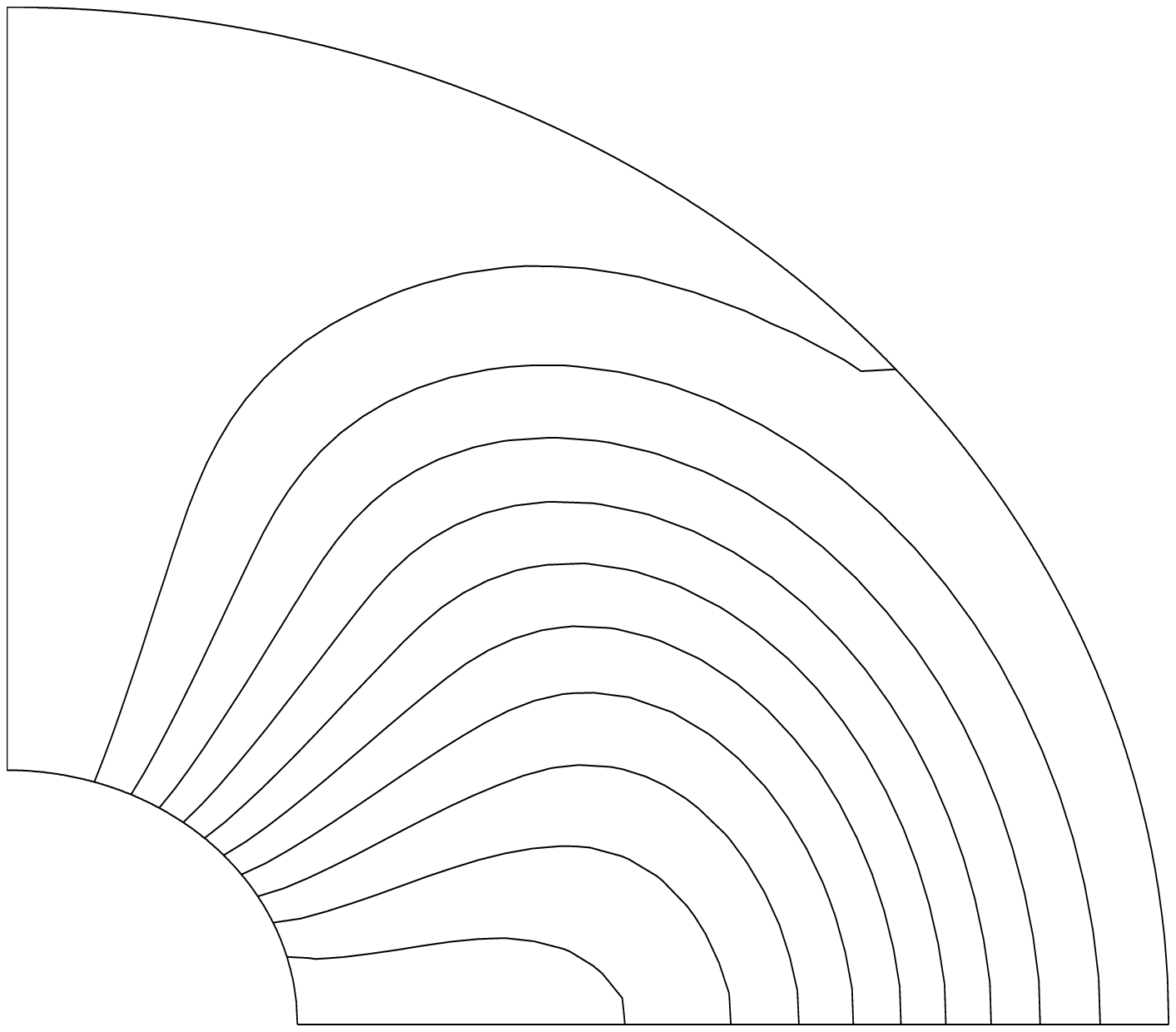,width=250pt}}}\end{center}
\caption[]{Field configuration at $t = 0.1$ starting from the initial configuration
shown in Fig.~3 evolved with $\eta = 1$.}
\label{fig4}
\eef

\bef
\begin{center}{\mbox{\epsfig{file=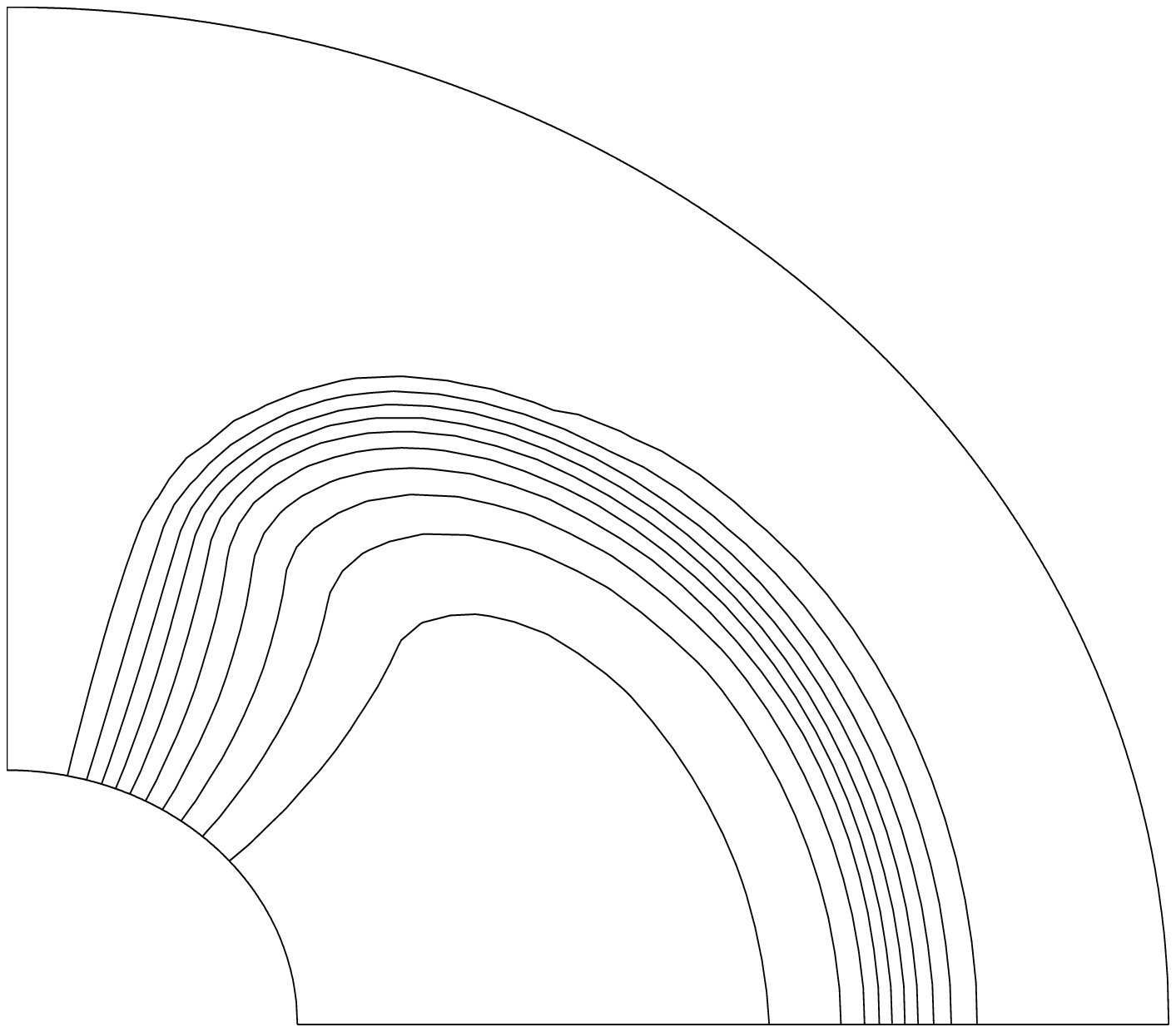,width=250pt}}}\end{center}
\caption[]{Field configuration at $t = 0.1$ starting from the initial configuration
shown in Fig.~3 evolved with $\eta = 0.01$.}
\label{fig5}
\eef

\bef
\begin{center}{\mbox{\epsfig{file=fig06.ps,width=225pt}}}\end{center}
\caption[]{Evolution of the mid-latitude surface field with time. The curves 1 and 2 correspond to the cases
of Fig.~\ref{fig4} ($\eta = 1)$ and Fig.~\ref{fig5} ($\eta = 0.01$) respectively.}
\label{fig6}
\eef

\bef
\begin{center}{\mbox{\epsfig{file=fig07.ps,width=225pt}}}\end{center}
\caption[]{Variation of $T_{1/2}$ for the surface field with $\eta$ when the initial field configuration is given by
Fig.~\ref{fig3}.}
\label{fig7}
\eef

\ni We have already mentioned that the magnetic fields in the uppermost molten layers just below the surface would be 
subject to magnetic buoyancy. Let us first present some calculations without including magnetic buoyancy. Now, to calculate 
the evolution of the field in time, we also need to specify the value of $\eta$. Figs.\ \ref{fig4} and \ref{fig5} show 
the magnetic field configuration after time 0.1 for $\eta = 1$ and $\eta = 10^{-2}$ respectively (in units in which $r_s$ and 
$K_3$ are taken as unity and we have also suppressed the physical dimensions of $\eta$ here).  When $\eta$ has the smaller 
value $10^{-2}$, the magnetic field is nearly frozen in the plasma and 
gets buried underneath the layer of accreted material, as seen in Fig.~\ref{fig5}. However, if $\eta$ has the larger value of 
1, then the field emerges out to the surface due to Ohmic diffusion and it is no longer possible for the accreting material 
to bury the magnetic field very effectively (Fig.~\ref{fig4}). The time evolution of $B_r$ at a point just below the surface 
at the mid-latitude is shown in Fig.~\ref{fig6}.  It is seen that $B_r$ decreases more rapidly if the diffusivity is low. At 
the first sight, this may appear counter-intuitive. We would normally expect a lower diffusivity to make a magnetic field decay 
less rapidly. Here the situation is reversed because the flow of accreting material is able to bury the magnetic field better if $\eta$ 
is lower and the magnetic field is frozen in the material more effectively. From a plot like Fig.~\ref{fig6}, one can
 easily estimate the time \Thalf\ in which the magnetic field near the surface falls to half of its initial value.  Figure~7 shows 
how \Thalf\ varies with $\eta$. When $\eta$ is large, the flow cannot carry the magnetic field with it and \Thalf\ is large.  
On the other hand, a smaller $\eta$ makes the magnetic field frozen in the plasma and the field can be carried away in the time 
scale of the flow.  Hence, for smaller values of $\eta$, \Thalf\ reaches an asymptotic value which is essentially the time scale 
of flow in the upper thin layer. The fact that \Thalf\ reaches an asymptotic value even before the value of $\eta$ is reduced 
to $10^{-2}$ is of some significance. Our code becomes numerically unstable if we make $\eta$ an order smaller than $10^{-2}$. 
On the basis of Fig.~\ref{fig7}, we expect that the behaviour of the system should not change on lowering the value of $\eta$ 
below $10^{-2}$, even though we are unable to run our code for such values.

\subsubsection{With Magnetic Buoyancy}

\bef
\begin{center}{\mbox{\epsfig{file=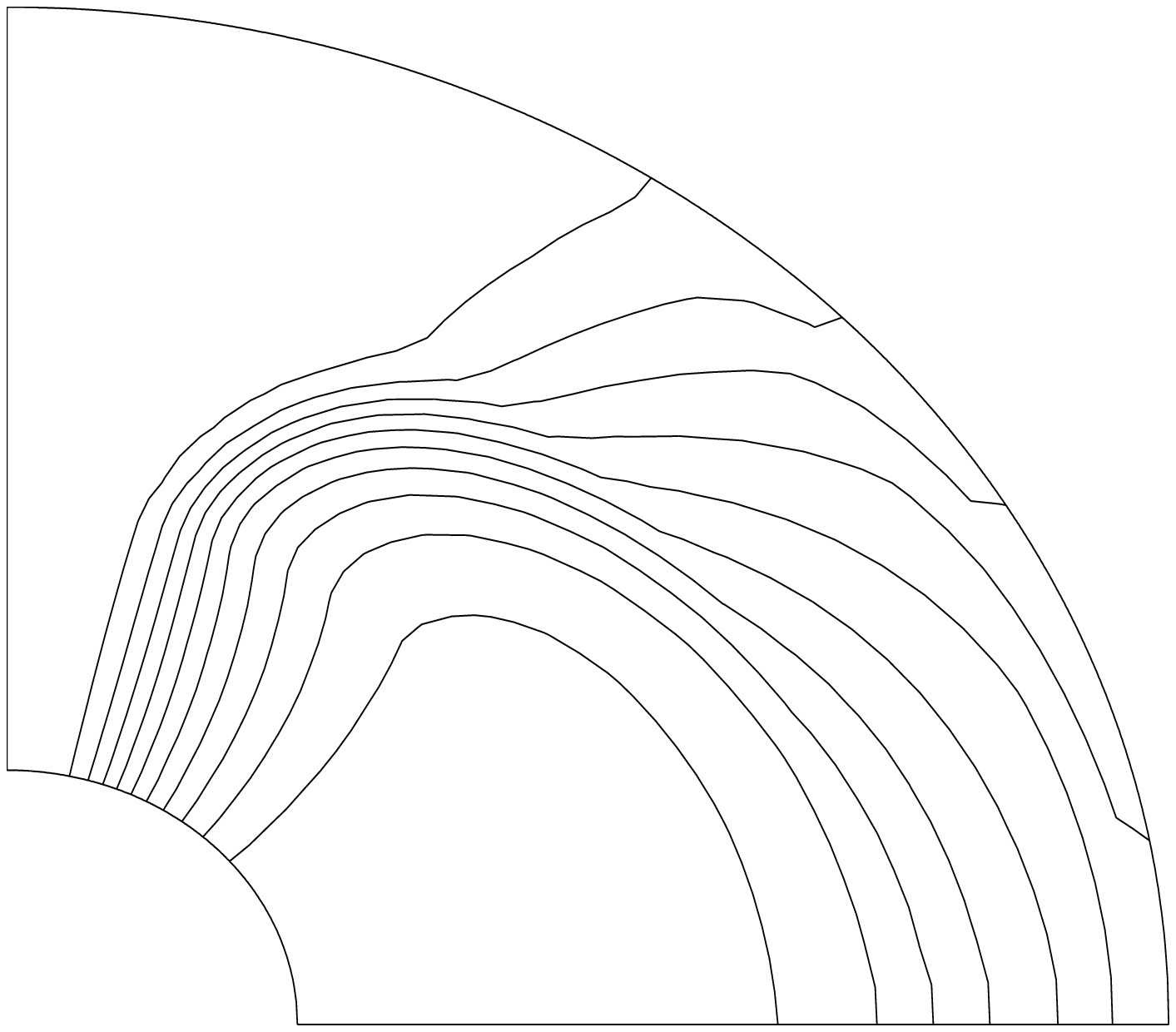,width=250pt}}}\end{center}
\caption[]{Field configuration at $t = 0.1$ starting from the initial configuration
shown in Fig.~3 evolved with $\eta = 0.01$ in the presence of magnetic buoyancy. Here, we
have taken $v_{\rm mb} = 50$.}
\label{fig8}
\eef

\ni In the uppermost layers of a neutron star where matter is in a molten state, magnetic buoyancy would make the magnetic 
field rise against gravity. It has been shown in the context of the solar convection zone that, when rotation is important, 
the magnetic field rises parallel to the rotation axis rather than radially outward (Choudhuri \& Gilman 1987, Choudhuri 1989).
In a rapidly spinning neutron star, the magnetic field may tend to rise parallel to the rotation axis. But, to simplify the problem, 
we are assuming that the magnetic field rises radially. The results of the present calculation are not expected to change much 
if we make the rise parallel to the rotation axis.  \\

\bef
\begin{center}{\mbox{\epsfig{file=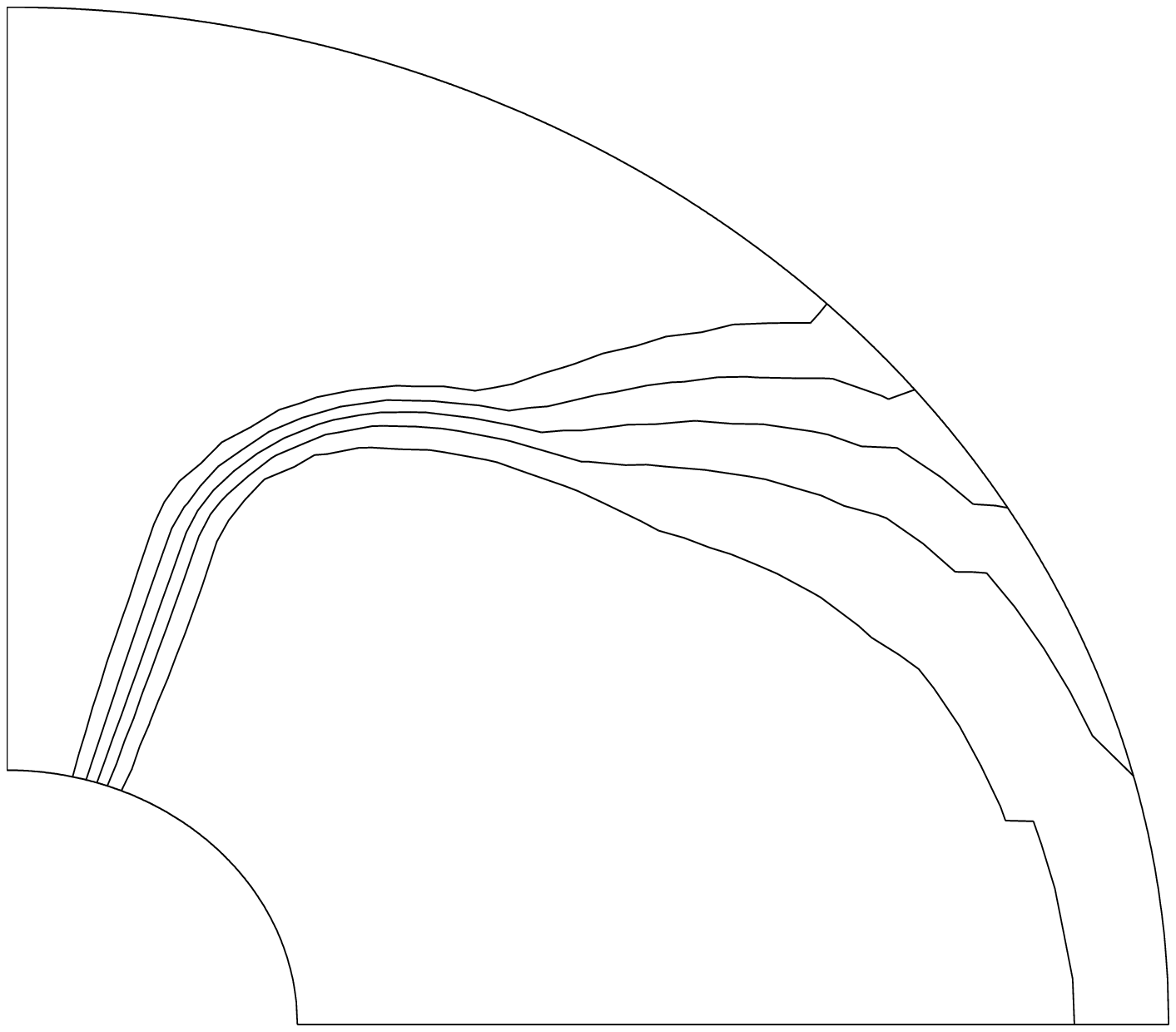,width=250pt}}}\end{center}
\caption[]{The field configuration of Fig.~\ref{fig8} evolved to $t = 0.5$.}
\label{fig9}
\eef

\bef
\begin{center}{\mbox{\epsfig{file=fig10.ps,width=225pt}}}\end{center}
\caption[]{Evolution of the mid-latitude surface field with time. The curves 1 and 2 are for cases without and with
magnetic buoyancy corresponding to Fig.~\ref{fig5} and Fig.~\ref{fig8} respectively. For both the cases, $\eta = 0.01$
and $v_{\rm mb} = 50$ for curve 2.}
\label{fig10}
\eef

\ni We capture the effect of magnetic buoyancy by superimposing on our velocity field given by eq.(\ref{vr1}-\ref{vth3}) a 
radial velocity in the uppermost layer. Hence, in the top layer $r_m < r <r_s$, we add a velocity given by
\beq
v_r = v_{\rm mb} (r - r_m).  
\eeq
The parameter $v_{\rm mb}$ determines the strength of magnetic buoyancy. It should be mentioned here that this is just a
simplified
way of modeling the effect of magnetic buoyancy. In reality, the upward velocity due to buoyancy must be proportional to the
square of the field strength at any point (Cumming, Zweibel \& Bildsten 2001). Since, we are only interested in obtaining the 
gross features of the possible screening
effect, here we work with the above mentioned model velocity instead of going into further details. We have found that our results 
remain virtually unaltered 
on taking $v_{\rm mb}$ equal to 50.0 or 100.0. This means that the effect of magnetic buoyancy saturates when $v_{\rm mb} = 50.0$ and 
things do not change any more on increasing $v_{\rm mb}$. We now present calculations with $v_{\rm mb} = 50.0$. We again begin with the 
initial configuration given by eq.(\ref{eq_dip}) and shown in Fig.~\ref{fig3}. Now, Fig.~\ref{fig8} shows the field at $t=0.1$
for $\eta = 10^{-2}$. The only difference between Fig.~\ref{fig5} and Fig.~\ref{fig8} is that Fig.~\ref{fig8} is obtained with 
magnetic buoyancy included. On comparing the two figures, we find that magnetic buoyancy in the uppermost layer has the effect of
making magnetic field lines radial there. In Fig.~\ref{fig5}, we have seen clearly that the equator-ward flow completely screens
the magnetic field from the uppermost surface layer. This has not happened in Fig.~\ref{fig8} where magnetic buoyancy has made 
the magnetic field emerge through the uppermost layer. To show how the magnetic field will keep evolving from the configuration 
of Fig.~\ref{fig8}, we show the magnetic configuration at a later time 0.5 in Fig.~\ref{fig9}. It is clear from Figures~\ref{fig8}
and \ref{fig9} that the slow flow in the deeper layers carries the magnetic field deeper down as well as to higher latitudes from 
the equatorial region. How the magnetic field $B_r$ at mid-latitude near the surface changes with time is shown in Fig.~\ref{fig10}.
For the sake of comparison, in this figure we again show the evolution of $B_r$ without magnetic buoyancy for $\eta = 10^{-2}$ 
(which was shown in Fig.~\ref{fig6} also). Although the magnetic field in the case with buoyancy has not been screened by the 
equator-ward flow in the upper layer, as in the case without buoyancy, we still find that $B_r$ keeps decreasing even in the case 
with buoyancy. The time in which $B_r$ decreases appreciably is of the order 0.1. We have already estimated in \S2 that the time 
scale for the slow flow in the interior is 0.5.  This clearly indicates that the decrease in $B_r$ near the surface is caused by 
the magnetic field being carried to deeper layers by the slow flow in the interior. Even though magnetic buoyancy is now allowing 
the magnetic field to emerge through the top layer and it is not possible for the equator-ward flow in that layer to screen the 
magnetic field, we conclude that the magnetic field still gets screened from the outside by the effect of the slow interior flow 
which pushes the magnetic field deeper down. In other words, the magnetic field gets buried in the time scale of the slow interior
flow when magnetic buoyancy is taken into account (instead of getting buried in the time scale of the faster flow of the top layer 
as in the case when buoyancy is not included). \\

\bef
\begin{center}{\mbox{\epsfig{file=fig11.ps,width=225pt}}}\end{center}
\caption[]{Variation of $T_{1/2}$ for the surface field with $\eta$ when the initial field configuration is given by
Fig.~\ref{fig3} and in presence of magnetic buoyancy ($v_{\rm mb} = 50$).}
\label{fig11}
\eef

\ni Letting the field evolve in time for different values of $\eta$, we can find \Thalf\ for the different values when magnetic 
buoyancy is included. Fig.~\ref{fig11} shows how \Thalf\ varies with $\eta$ in this case. As in Fig.~\ref{fig7}, we find that 
\Thalf\ becomes independent of $\eta$ when $\eta$ is sufficiently small to make the magnetic field nearly frozen. The asymptotic 
value of \Thalf\ in Fig.~\ref{fig11}, however, is much larger than what it is in Fig.~\ref{fig7}. The asymptotic value of \Thalf\ 
in Fig.~\ref{fig7} basically corresponds to the time scale of the equator-ward flow in the top layer, whereas the asymptotic value 
in Fig.~\ref{fig11} corresponds to the time scale of the slower flow in the interior.  

\subsection{Crustal Field Configuration}

\bef
\begin{center}{\mbox{\epsfig{file=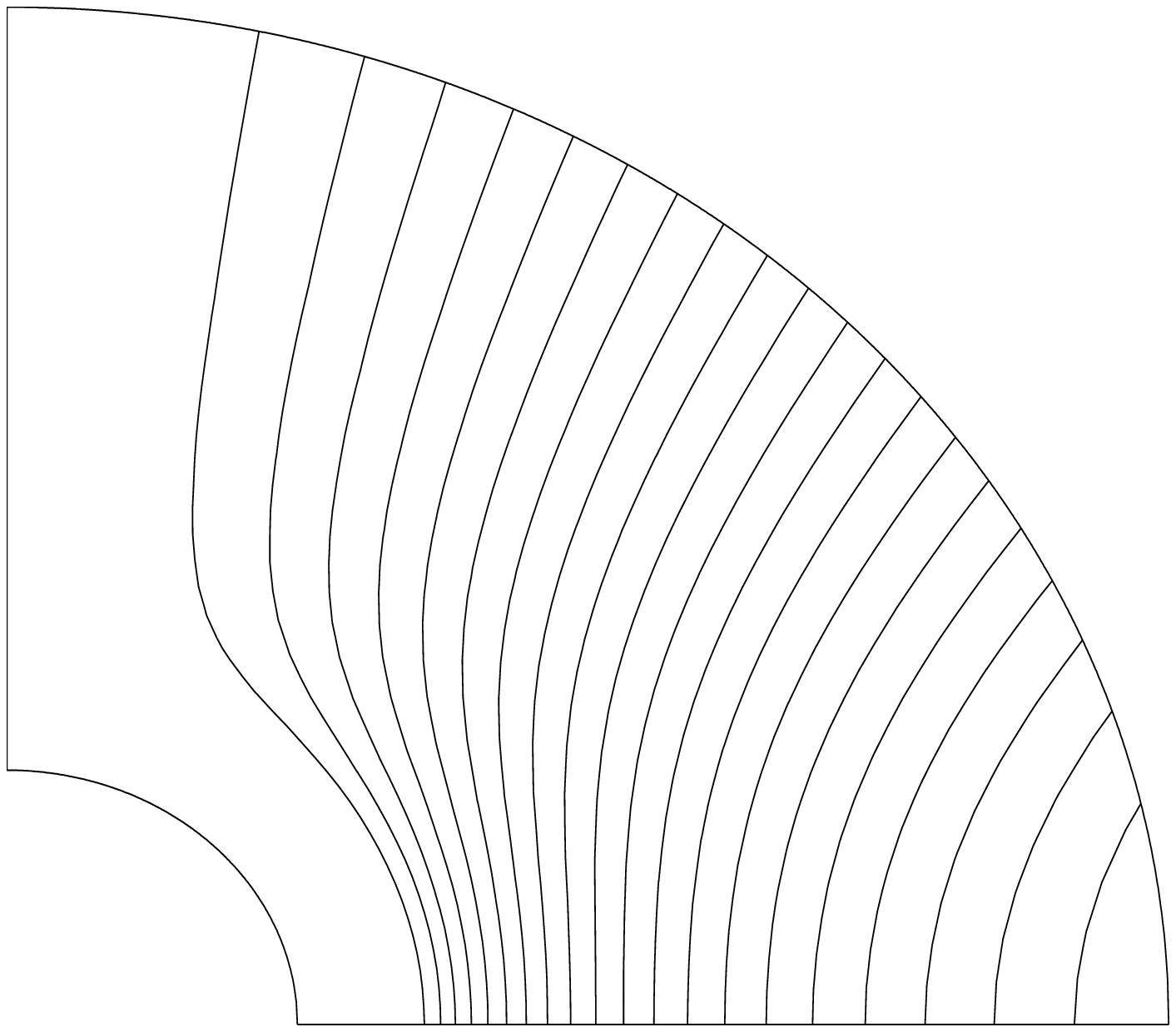,width=250pt}}}\end{center}
\caption[]{The initial configuration of the crustal field, in a region 
defined by $0.25 \leq r \leq 1$ 
and $0 \leq \theta \leq \pi/2$.}
\label{fig12}
\eef

\ni We now show what happens if the magnetic field is confined to the crust. For this purpose, we begin with the magnetic 
configuration shown in Fig.~\ref{fig12}. Then Figures~\ref{fig13} and \ref{fig14} respectively show the field configurations
at $t= 0.1$ without and with magnetic buoyancy (using $\eta = 10^{-2}$). It is again found that the flow of accreted material 
in the top layer quickly screens the magnetic field in the absence of magnetic buoyancy, whereas on inclusion of magnetic 
buoyancy we find that the field is able to emerge through the top layer. The interior flow carries the magnetic field slowly to 
deeper layers and to higher latitudes. It is interesting to note that the magnetic field eventually tends to take up 
configurations which do not look very different from the configurations which arise if we begin with an initial magnetic field 
passing through the core (compare Figures~\ref{fig13}--\ref{fig14} with Figures~\ref{fig5} and \ref{fig8}). We thus expect the 
results to be qualitatively similar even if we start from quite different initial configurations. \\

\bef
\begin{center}{\mbox{\epsfig{file=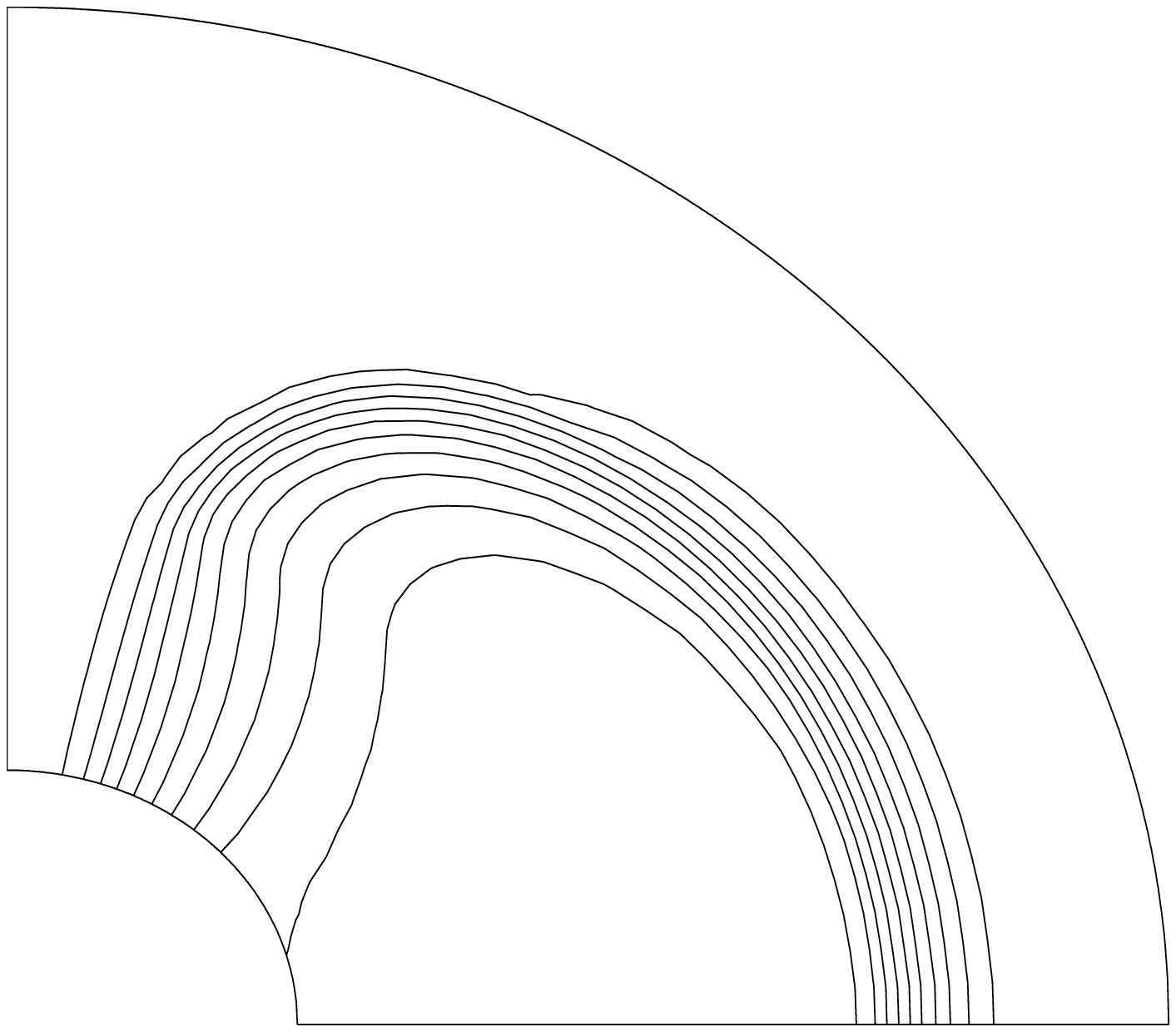,width=250pt}}}\end{center}
\caption[]{Field configuration at time $t = 0.1$ starting from the initial
configuration shown in Fig.~\ref{fig12} evolved
with $\eta = 0.01$ and without magnetic buoyancy.}
\label{fig13}
\eef

\bef
\begin{center}{\mbox{\epsfig{file=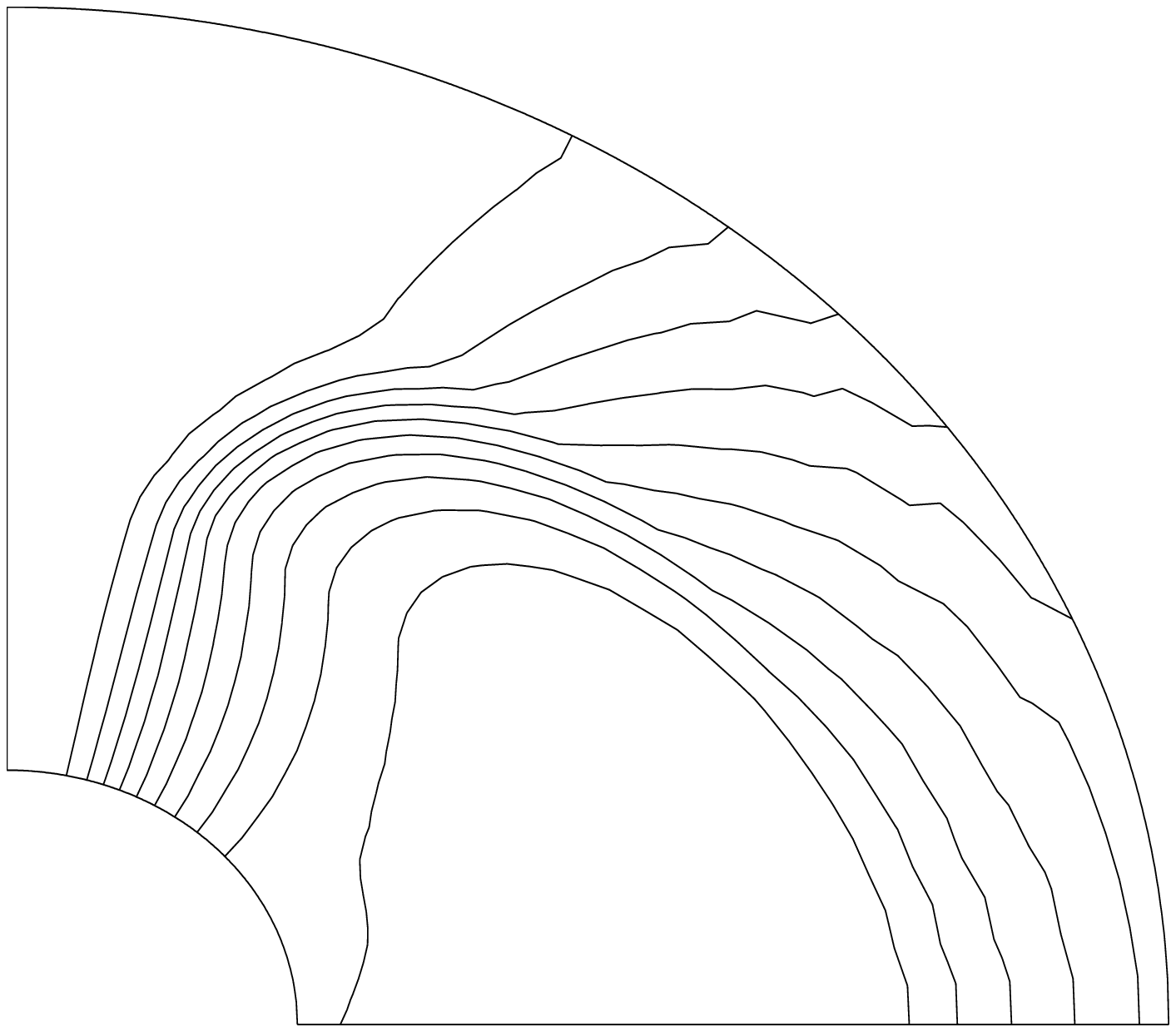,width=250pt}}}\end{center}
\caption[]{Field configuration at time $t = 0.1$ starting from the initial
configuration shown in Fig.~\ref{fig12} evolved
with $\eta = 0.01$ and with magnetic buoyancy ($v_{\rm mb} = 50$).}
\label{fig14}
\eef

\ni To see the effect of the velocity field very clearly, we also study what happens if the velocity field is switched off 
and the magnetic field is allowed to decay due to diffusion alone. Fig.~\ref{fig15} shows what the magnetic field starting 
from the initial configuration of Fig.~\ref{fig12} would look at $t=1.0$ if the velocity field is switched off. Since 
diffusion is a slower process for the values of parameters we are using, note that we have evolved the magnetic field for a 
longer interval compared to what was done in the other figures. We see that the magnetic field has spread outside the region 
where it was initially confined. We point out that such a study of pure diffusion would not be possible with the initial
potential field passing through the core (shown in Fig.~\ref{fig3}) that we considered in the earlier subsections. Our 
bottom boundary condition specified in the Appendix is such that $A$ would not change at the bottom if the velocity field were switched off. Hence, if 
we start from the initial potential field (given by eq.(\ref{eq_dip})) and allow it to evolve under diffusion alone, it is 
easy to see that the field remains invariant. \\ 

\bef
\begin{center}{\mbox{\epsfig{file=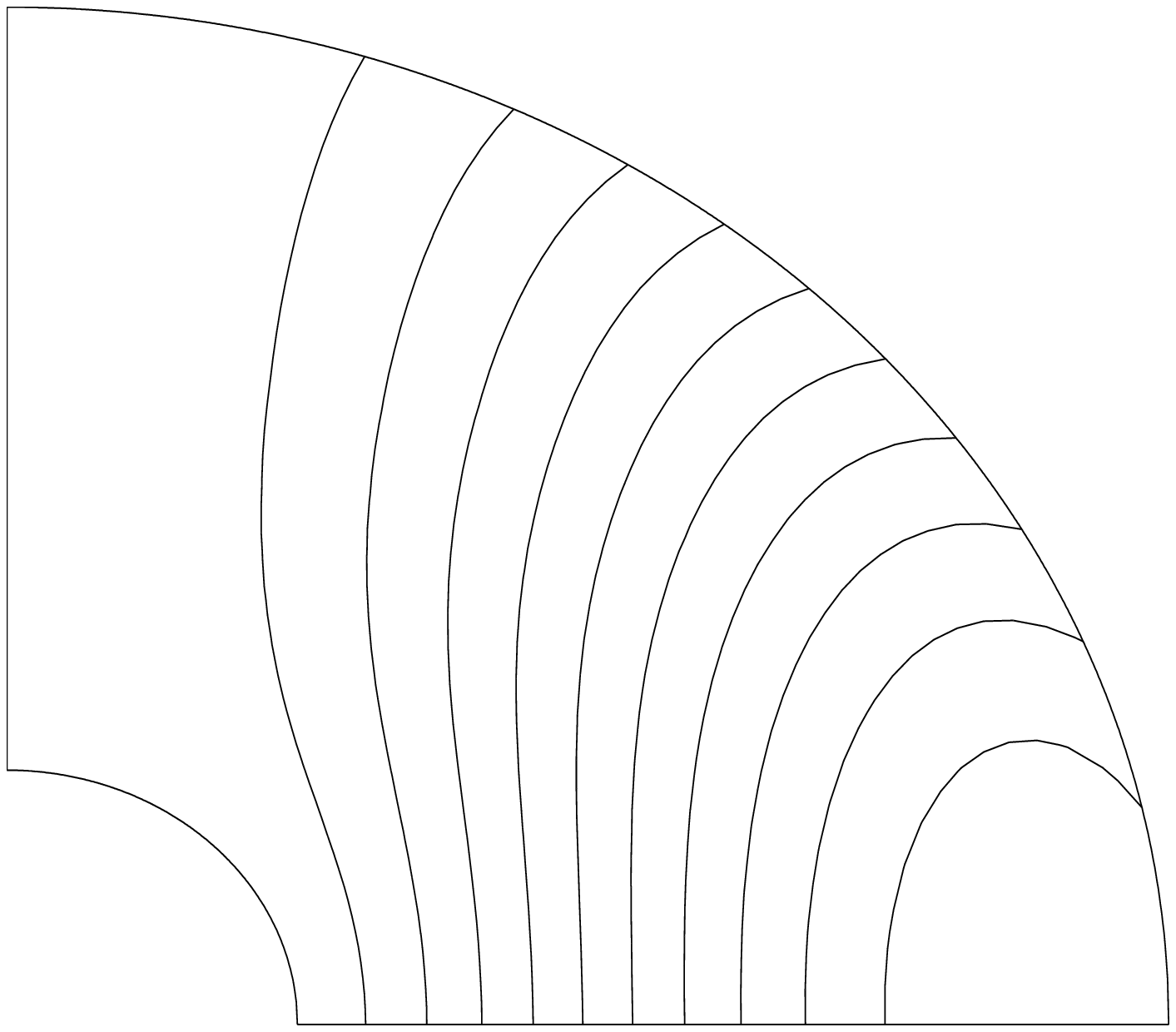,width=250pt}}}\end{center}
\caption[]{Field configuration at time $t = 1.0$ starting from the initial
configuration shown in Fig.~\ref{fig12}, 
in the case of pure diffusion, with $\eta = 0.01$.}
\label{fig15}
\eef

\bef
\begin{center}{\mbox{\epsfig{file=fig16.ps,width=200pt}}}\end{center}
\caption[]{Evolution of the mid-latitude surface field with time. The curves 1, 2 and 3 correspond to the cases
seen in Fig.~\ref{fig13}, Fig.~\ref{fig14} and Fig.~\ref{fig15} respectively.}
\label{fig16}
\eef

\ni Fig.~\ref{fig16} shows how $B_r$ at mid-latitude near the surface evolves for the three cases shown in
 Figures~\ref{fig13}-\ref{fig15} (i.e.\ without magnetic buoyancy, with magnetic buoyancy and without any flow). We see that 
pure diffusion has a larger time scale. If there were no magnetic buoyancy, then the magnetic field would have been screened 
in a very short time scale arising from the flow of accreted material in the top layer. On account of magnetic buoyancy, we 
obtain an intermediate time scale between these two, which is essentially the time scale of the slow interior flow.

\subsection{Field Confined to a Narrow Layer}

\ni We now argue that our general conclusions do not change on reducing the thickness of the layer in which the flow takes place. 
We point out in the next section that we expect $r_m = 0.99$ in a neutron star (we have been taking $r_m = 0.75$ in the above 
exploratory calculations). We find that our code does not retain numerical stability if $r_m$ is increased to 0.99. However, we 
have been able to obtain results by increasing the value of $r_m$ up to 0.95 and find that our general conclusions inferred
in the earlier subsections do not change. We now present our results obtained by taking $r_m = 0.95$ and $r_b = 0.85$. The
initial field configuration is shown in Fig.~\ref{fig17}. We also change the value of $\alpha$ in eq.(\ref{eq_rho}) to 20 to 
simulate a very sharp drop in the density profile. \\

\ni We find that in absence and in presence of magnetic buoyancy the field configuration gets modified to the ones seen in
Fig.~\ref{fig18} and Fig.~\ref{fig19} respectively. It is evident that the generic nature of field evolution is very much the
same as that seen in the previous subsection.  In the earlier case, however, magnetic buoyancy
saturated at $v_{\rm mb} = 50$ and the nature of field 
evolution did not change on increasing $v_{\rm mb}$ beyond. In the present case, 
the similar effect is seen around $v_{\rm mb}=500$.  This difference is not difficult to understand if
we keep in mind that $v_{\rm mb}$ has to be multiplied by $(r-r_m)$ to get the buoyant rise speed, as
seen in (23). Whereas the maximum value of $(r-r_m)$ was 0.25 for the calculations in the previous
subsections, here its maximum value is 0.05.  To compensate for this, we need to take a larger $v_{\rm mb}$
to get buoyant rise speeds of similar magnitude. \\

\ni For comparison, we also show the case of pure diffusion in Fig.~\ref{fig20}. To compare the time scales of evolution, in 
Fig.~\ref{fig21} we have plotted the decay of the surface field (near mid-latitude) for the three cases - a) without magnetic
buoyancy, b) with magnetic buoyancy and c) for pure diffusion. Conforming to our expectation (which has already been seen in
the previous sub-section), the time scales of field decay without buoyancy, with buoyancy and that of pure diffusion arrange
themselves in an increasing order. Therefore, in conclusion we can say that all the characteristic features of our calculation
are also borne out by confining the physics to a very narrow layer near the surface. This vindicates the attempt of generalizing
our results to the case of neutron stars (which we describe in some detail in the next section), even though the exact calculations
could not be performed by our numerical code.

\bef
\begin{center}{\mbox{\epsfig{file=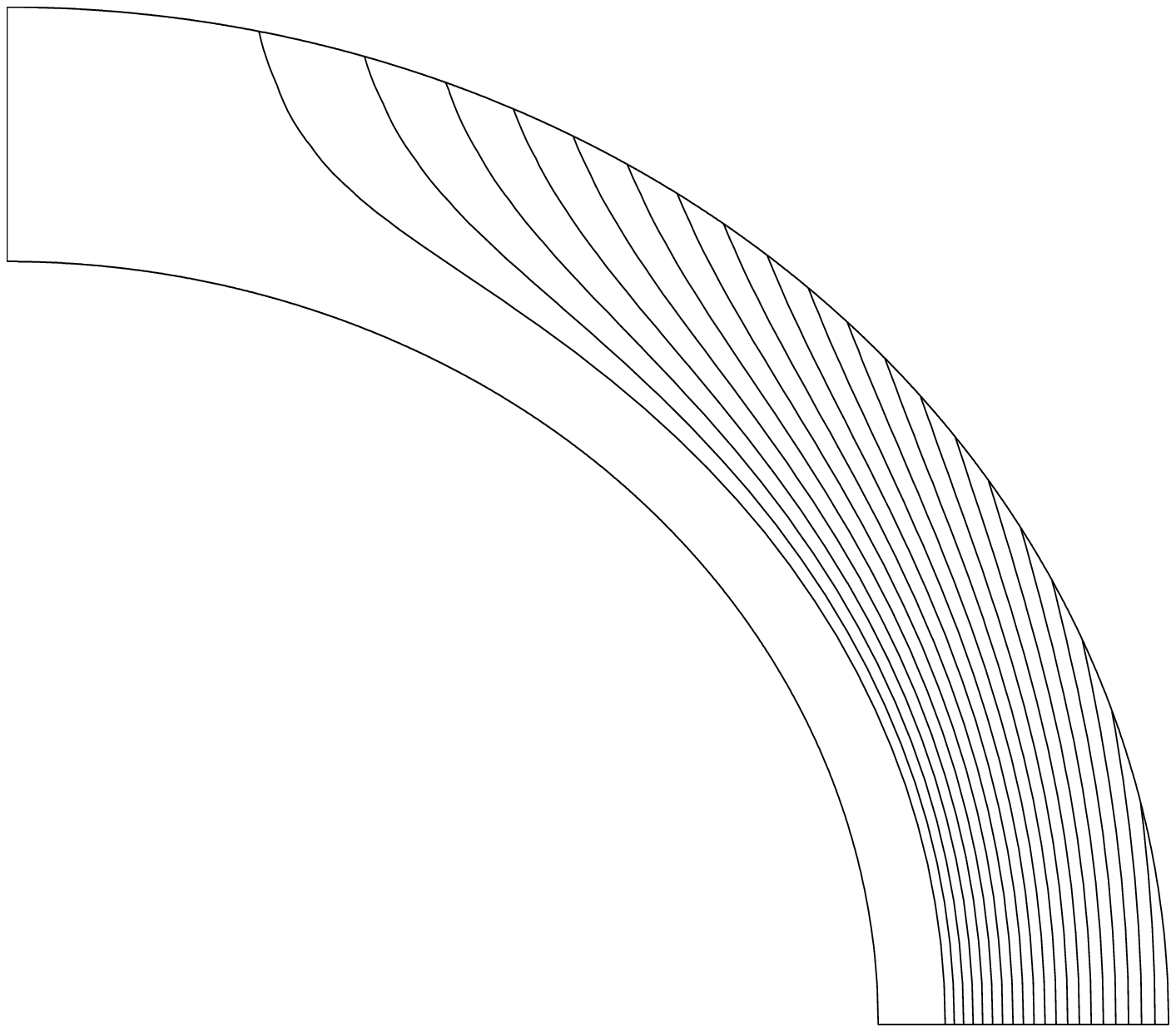,width=250pt}}}\end{center}
\caption[]{Initial field configuration similar to that in Fig.~\ref{fig12} but 
confined to a narrower region with 
$r \leq 0.8$.}
\label{fig17}
\eef

\bef
\begin{center}{\mbox{\epsfig{file=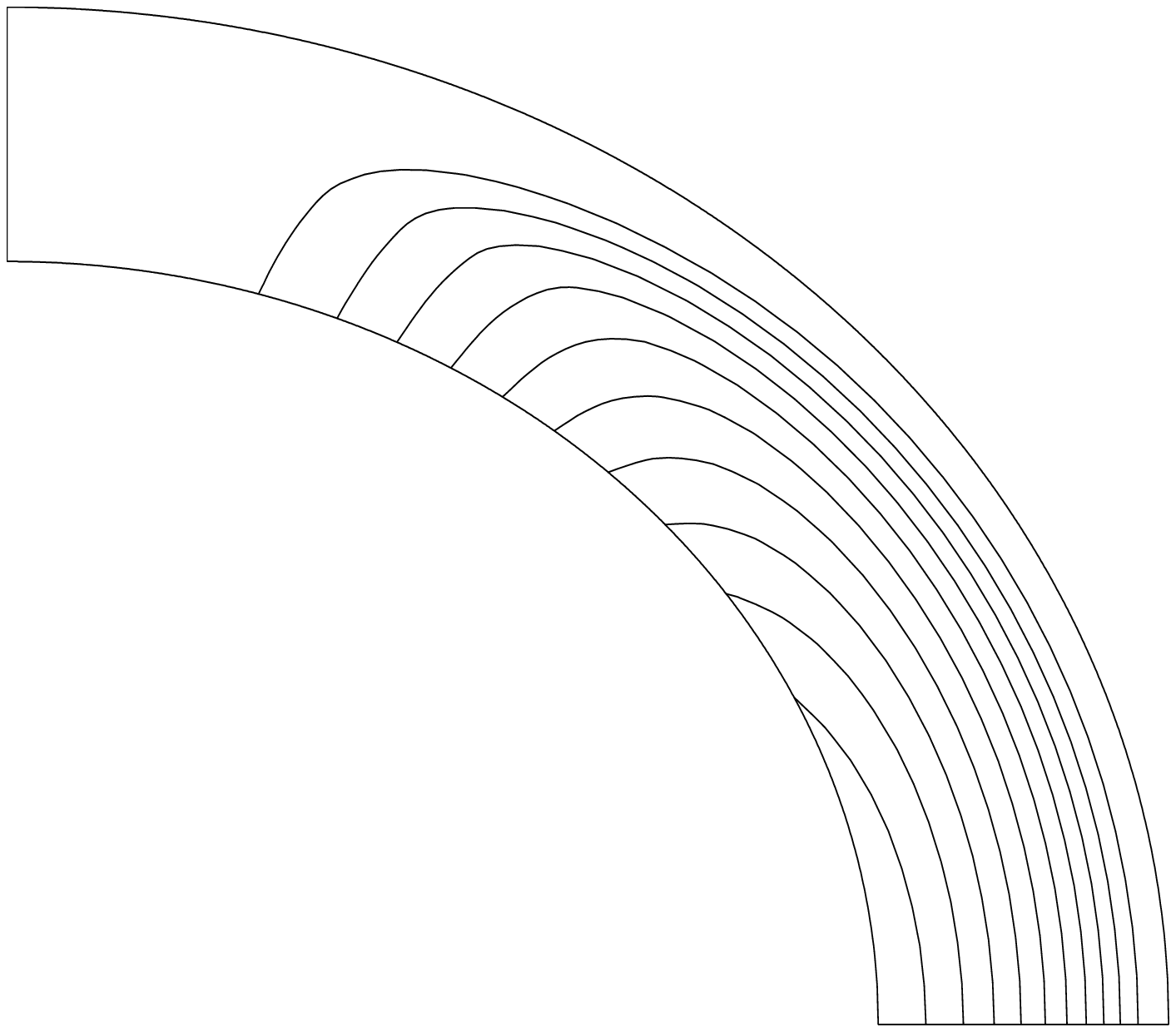,width=250pt}}}\end{center} 
\caption[]{Field configuration at time $t = 0.1$ starting from the initial
configuration shown in Fig.~\ref{fig17} evolved
with $\eta = 0.1$ and without magnetic buoyancy.}
\label{fig18}
\eef

\bef
\begin{center}{\mbox{\epsfig{file=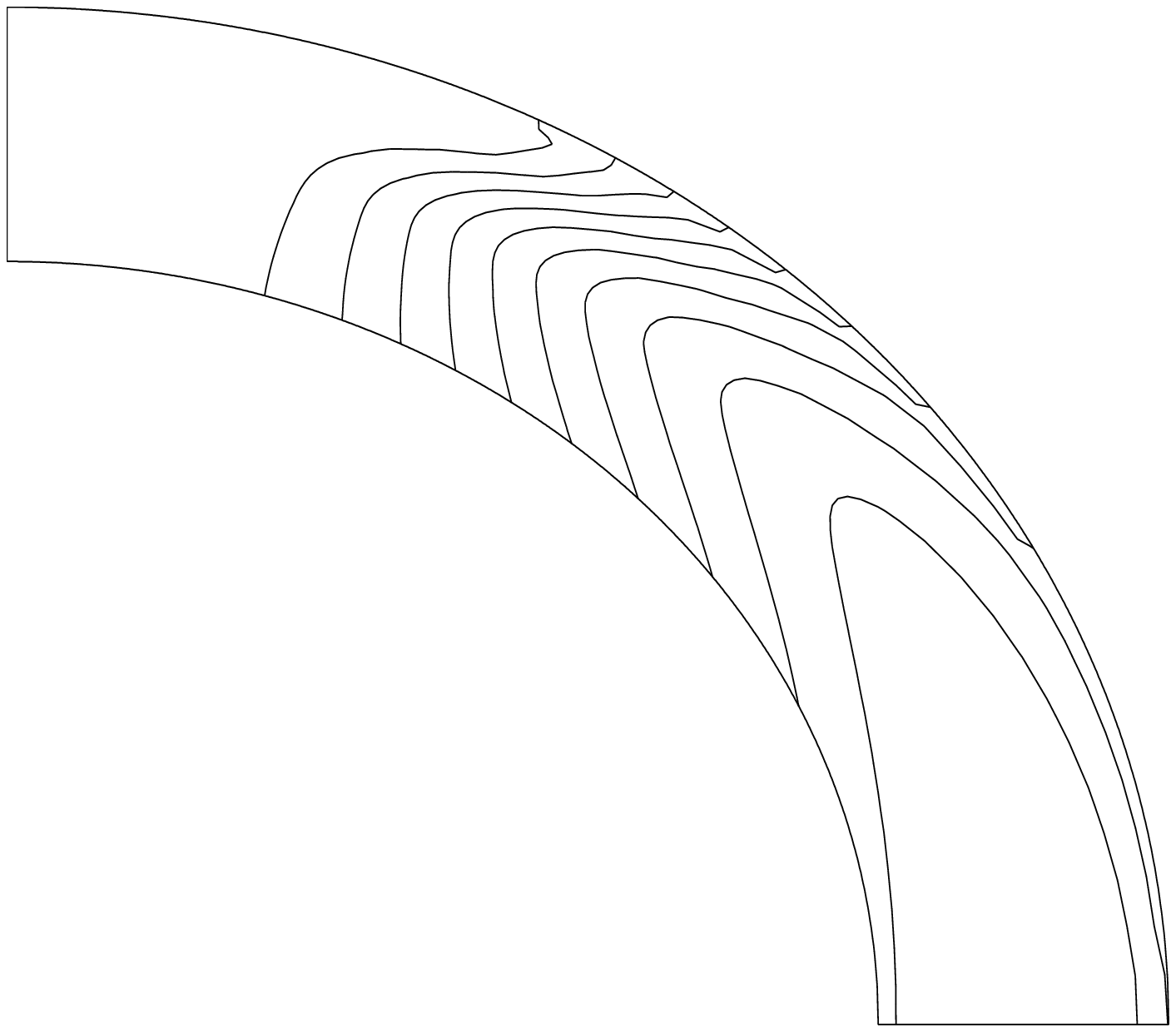,width=250pt}}}\end{center}
\caption[]{Field configuration at time $t = 0.1$ starting from the initial
configuration shown in Fig.~\ref{fig17} evolved
with $\eta = 0.1$ and with magnetic buoyancy ($v_{\rm mb} = 500)$.}
\label{fig19}
\eef

\bef
\begin{center}{\mbox{\epsfig{file=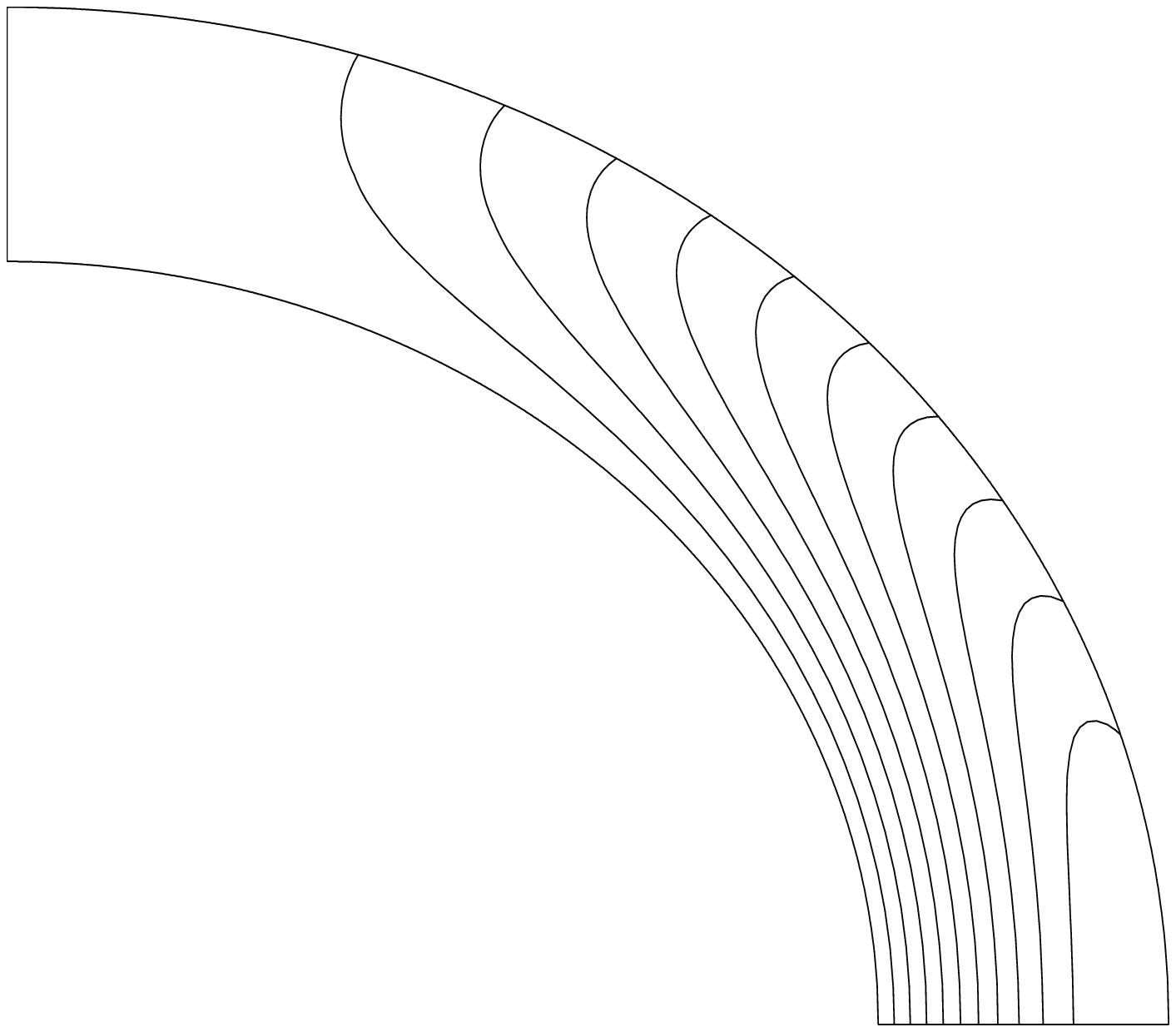,width=250pt}}}\end{center}
\caption[]{Field configuration at time $t = 0.5$ starting from the initial
configuration shown in Fig.~\ref{fig17}, in the case of pure diffusion, 
with $\eta = 0.1$.}
\label{fig20}
\eef

\bef
\begin{center}{\mbox{\epsfig{file=fig21.ps,width=225pt}}}\end{center}
\caption[]{Evolution of the mid-latitude surface field with time. The curves 1, 2 and 3 correspond to the cases
seen in Fig.~\ref{fig18}, Fig.~\ref{fig19} and Fig.~\ref{fig20} respectively.}
\label{fig21}
\eef

\section{field evolution in neutron stars}

\ni In a real neutron star, the equator-ward flow must be confined to a very narrow layer near the surface within which density 
varies by several orders of magnitude. Our numerical code is unable to handle such a severe density contrast. However, from the 
results presented in the previous section, the following important parameter-independent conclusions emerge :
\ben
\i If the flow time scale is shorter than the diffusion time scale, then it is possible for the flow to carry the magnetic field 
with it and eventually bury the field under the surface.
\i In the absence of magnetic buoyancy, the magnetic field gets buried in the time scale of the equator-ward flow in the top layer.
\i When magnetic buoyancy is taken into account, the magnetic field is buried in the time scale of the much slower counter-flow 
under the top layer.
\een
These results seem fairly robust and we expect them to be valid in the case of neutron stars too. Therefore, we are able to 
make some inferences about the neutron stars on the basis of our estimates of the time scale of diffusion, the time scale of 
equator-ward flow of accreted matter and the time scale of very slow motion of the solid material under the top layer. We now 
attempt to estimate these various time scales. \\
 
\ni The magnetic field of the isolated neutron stars is typically $\sim 10^{12}$~G, which makes the magnetic pressure at the polar 
cap $\sim 10^{23}$~dyne cm$^{-2}$. As material starts collecting in the polar regions, excess pressure builds up, and when this 
excess pressure becomes $\sim 10^2$ of the magnetic pressure, the accumulated material flows sidewise overcoming the magnetic 
stresses (Brown \& Bildsten 1998). From the equation of state of the neutron star material, we find that the density at which the 
pressure becomes $\sim 10^{25}$~dyne cm$^{-2}$ is $\sim 10^9~\gcc$. From the density profile of a 1.4~\msun neutron star, assuming 
a moderately stiff equation of state, we find that $\rho = 10^9~\gcc$ corresponds to a depth of 100 m below the surface, where $r$ 
has the approximate value $\sim 0.99 r_s$ ($r_s$ is radius of the star, which turns out to be about 10 km). (For a discussion on 
the equation of state used here and the calculation of the density profile of a 1.4~\msun neutron star, see Konar 1997). We expect 
that the excess pressure of accreting material at the poles will induce equator-ward flow up to this depth of 100 m. \\

\ni Let us assume that the rate of material accretion in the polar region is \mdot. If the polar cap has a circumference of 
$2 \pi l$ and there is an outflow across this circumference with velocity $v$ and up to depth $h$, then we must have:
\beq
\dot{M} \approx 2 \pi l h \rho v \,.
\eeq
The Eddington accretion rate for a neutron star is $\sim 10^{-8}$ \msun\ yr$^{-1}$, i.e.\ about $10^{18}$ gm s$^{-1}$ in cgs units.  
Taking $l \approx 1$ km, $h \approx 100$ m, $\rho \approx 10^9$ gm cm$^{-3}$, we find the velocity of outflow to be:
\beq
v \approx 10^{-1} \mbox{cm s}^{-1} \,.
\eeq
Combining this velocity with a length of 10 km (i.e. the typical radius of a neutron star), we get a time scale of about 1 year 
for the equator-ward flow of accreted matter. \\

\ni We now estimate the velocity of the slow counter-flow of solid material under the surface. Assuming a density increase of 
100, we found that the velocity of counter-flow in the inner layers is about 100 times less than the velocity of equator-ward flow 
in the top layer (see Fig.~\ref{fig2} and discussions in the text about it). We have estimated the density in the layer of 
equator-ward flow to be about $10^9$ gm cm$^{-3}$, whereas the density some distance below the surface is $\sim 10^{14}~\gcc$.
An increase of density by a factor of $10^5$ would lead to a decrease of velocity by the same factor. Although geometrical factors 
such as the radial depth of the region across which the counter-flow takes place would have some role, here we are interested only
in rough orders of magnitude. Dividing the velocity at the surface by this factor of $10^5$, we get a velocity of about 
$10^{-6}$~cm s$^{-1}$. This corresponds to a time scale of $\sim 10^5$ years. We can conclude that the magnetic field of the 
accreting neutron star would get screened in 1 year in the absence of magnetic buoyancy and in about $10^5$ years if magnetic 
buoyancy is important, provided the diffusion time scale is larger than these time scales and the magnetic field can be assumed 
as nearly frozen in the material. \\

\ni The electrical conductivity $\sigma$ in the interior of an accretion heated neutron star is $\sim 10^{26}$ s$^{-1}$ so that 
$\eta \sim 10^{-6}$ cm$^2$ s$^{-1}$. Using a length scale of about 1 km, we obtain a diffusion time scale $\sim 10^{16}$~s or 
$\sim 10^{9}$ years. It is true that the electrical conductivity of the crust can be as low as $\sim 10^{20}~s^{-1}$ in the 
surface layers and the diffusive time scale calculated on the basis of that will be smaller by several orders. However, if 
magnetic buoyancy is dominant in the outer layers, the magnetic field comes out to the surface in any case and we need not 
bother about the role of Ohmic diffusion in the crust in making the magnetic field emerge through the accreted matter. The 
important point to note here is that the diffusion time scale in the interior is much larger than the estimated time scale of 
$10^5$ years of the slow flow in the interior. We, therefore, expect that the slow displacement of the solid material in the 
interior to be able to carry the magnetic field and bury it in the time scale of $10^5$ years. It should be mentioned here that 
this time scale happens to be of the similar order as the mass accretion time scale of most massive X-ray binaries. \\

\ni Let us now make a few comments on magnetic buoyancy. It is well known that magnetic buoyancy is particularly destabilizing 
in a region of super-adiabatic gradient, whereas its effect can be reduced in a region of sub-adiabatic gradient (Parker 1979, 
Moreno-Insertis 1983).  Although the top layer of the neutron star is expected to be sub-adiabatic, we believe that magnetic
buoyancy will still be important there.  In a sub-adiabatic region, a magnetic flux tube continues to rise as it exchanges heat 
with the surrounding medium and comes to thermal equilibrium (Parker 1979, \S8.8).  The very high thermal conductivity in the 
crust of the neutron star should keep magnetic buoyancy going even in a sub-adiabatic environment. So it is unlikely that the 
magnetic field of the neutron star can get screened in the time scale of 1 year of the flow of accreting matter on the surface.  
Even if magnetic buoyancy were suppressed for some unknown reason, the much shorter Ohmic diffusion time in the crust would make 
the magnetic field emerge out of the accreted matter (Cumming, Zweibel \& Bildsten 2001). However, it appears that the slow 
motion of the solid material in the interior of the neutron star may carry the magnetic field with it and eventually bury the 
magnetic field in its time scale of $10^5$ years.  It is true that we have not been able to present detailed calculations
appropriate for the neutron star and some of our arguments are heuristic. However, we humbly submit that this proposal is worth 
considering.

\section{Conclusion}

In the past there has existed a controversy regarding the possible mechanism of screening of the magnetic field of the neutron star 
by the accreting matter. We have tried to address the question with a 2-D simulation.  Although we could run our code only for 
thicknesses of flow layers much more than what they would be in a realistic situation, we have arrived at a few generic results 
which seem robust and independent of the thickness of flow layer.  So we believe that these results can be extrapolated to neutron 
stars. Basically we find that the magnetic field would get screened in the short time scale of the flow of accreting material in the 
top layer if magnetic buoyancy is neglected, whereas on inclusion of magnetic buoyancy the screening takes place in the somewhat 
longer time scale of the slow interior flows.  Since magnetic buoyancy is expected to be important for neutron stars, we believe 
that the second time scale of interior flows will be the appropriate time scale for screening the magnetic field of neutron stars.  
We have estimated this time scale to be of order $10^5$ years.\\

\ni It is remarkable that this is comparable to the time scale of accretion in massive X-ray binaries.  It must be an extraordinary 
coincidence that these two time scales arising out of completely different considerations turn out to be fairly close. This 
coincidence has important implications.  If the time scale for screening was much larger than the time scale for accretion, then the 
magnetic field could not be screened much during the accretion phase.  On the other hand, if the screening time scale happened to be 
much shorter than accretion time scale, then the magnetic field might have been screened so well that there would be virtually no 
magnetic field seen from outside.  Observationally we find that the magnetic fields of binary/millisecond pulsars are considerably 
weaker compared to magnetic fields of normal pulsars, but are not altogether negligible.  Can we take this as an observational
confirmation that the screening time scale should be of the similar order as the accretion time scale? \\

\ni Although considerations of neutron stars motivated us to undertake these calculations, we point out that what we have studied 
is the basic physics of magnetic field evolution in a situation of polar accretion. Our results should apply to other situations 
such as accretion onto white dwarfs. In the case of neutron stars, we have observations which seem to indicate that accreting
matter may screen the magnetic field.  We do not have observations to tell us what happens to magnetic fields in other stellar 
systems as a result of mass accretion. But the basic physics must be the same.\\
 
\ni Our model is based on certain assumptions. We have assumed a specific form of the velocity field, which essentially implies 
that the accreting matter primarily sinks near the equator where flows from the two different hemispheres meet and the matter in 
the underlying layers at the equator is thereby pushed downward as well as to higher latitudes in the form of a counter-flow.  
The other crucial assumption is that the dense matter in the lower layers is displaced very slowly while remaining solid and hence
the magnetic field remains embedded in the solid matter of interior flow regions where magnetic buoyancy is suppressed. Once we
make these assumptions, our results follow from fairly straightforward calculations.  Are our assumptions justified? We are not 
aware of any arguments that can be advanced against these assumptions and they do appear reasonable to us. To substantiate these 
assumptions fully, one will require some sophisticated fluid mechanics and sophisticated condensed matter physics of very dense
material. We hope that researchers more conversant in these subjects than us will look into these questions in future.  \\

\ni This being the first calculation of this type, we have restricted ourselves to a simple exploratory model. We could not run 
our code reducing the thickness of the upper layer to realistic values.  Researchers more competent in handling numerical 
instabilities should be able to do better than us. Also, we have made the simplifying assumption that our velocity field remains 
independent of time.  As magnetic field is pushed away, the polar cap widens and the accretion should take place over a broader 
region around the pole. We do not expect this to change our basic results, which seem quite robust. However, one will have to use 
a time-varying velocity field to build more realistic models of diamagnetic screening. All our calculations are based on a velocity 
field for which we could write down a convenient analytical expression. Since we can only make intelligent guesses as to what the 
velocity field may be like at the surface of the neutron star, it would be worthwhile to investigate other forms of the velocity 
field and to ascertain whether certain velocity fields are more efficient in diamagnetic screening than others. We are right now in
in the process of studying the screening of the magnetic field by other types of velocity fields, especially by velocity fields 
which change with time as the polar cap widens.  We hope to present our results in a future paper (Konar \& Choudhuri, in 
preparation). We end with the comment that our first exploratory study appears promising and the scenario presented by us should be 
investigated further.
  
\section*{Acknowledgments}
We thank Dipankar Bhattacharya for suggesting this problem, and Partha Sarathi Joarder and Dibyendu Nandy for help with the numerical 
code.  We are indebted to Mousumi Dikpati, whose code developed in the context of solar MHD provided the template for the development 
of the code used in the present calculations. A few subroutines of the older code are used in our programme. We also gratefully
acknowledge U.~R.~M.~E. Geppert's very detailed comments on our manuscript and the referee comments by Denis Konenkov - both of which
have helped to improve the paper significantly. ARC acknowledges the warm hospitality received at IUCAA Pune, where a part of this 
work was done while ARC was visiting as a Senior Associate.
\def\lr{\overline{L}_r}
\def\lt{\overline{L}_\theta}

\section*{Appendix}

\ni The basic features of our numerical code developed to solve eq.(\ref{eq_dadt}) are outlined here. The basic equation we have to solve is (3), which can be written
down in the form
\beq
\frac{\partial A}{\partial t} = L_r A + L_\theta A\,,
\eeq
where the operators $L_r$ and $L_\theta$ are given by
\ber
L_r A &=& - \frac{v_r}{r} \frac{\pa}{\pa r} (r A) \nonumber \\
      && + \eta \left[ \frac{\pa^2 A}{\pa r^2} + \frac{2}{r} \frac{\pa A}{\pa r} - \frac{A}{2 r^2 \sin^2 \theta} \right] \,, \\
L_\theta A &=& - \frac{v_\theta}{r \sin \theta} \frac{\pa}{\pa \theta} (\sin \theta A) \nonumber \\
           && + \eta \left[\frac{1}{r^2} \frac{\pa ^2 A}{\pa \theta^2} 
              + \frac{\cot \theta}{r^2} \frac{\pa A}{\pa \theta} - \frac{A}{2 r^2 \sin^2 \theta} \right] \,.
\eer
To solve this two-dimensional problem numerically, we employ the Alternating Direction Implicit (ADI)
scheme of differencing (see, for example, Ames 1977; Press {\it et al.\ } 1988). In this scheme, each time-step is divided in two halves. In the first half time-step
one direction (say, $r$) is advanced implicitly and the other direction (say, $\theta$) is advanced explicitly.
In the next half time-step the two directions are treated in the reverse manner.
If $A_{i,j}^m$ is the value of $A$ at the grid point $(i,j)$ at time step $m$,
then this implies the following two time steps:
\beq
A_{i,j}^{m+ \half} - A_{i,j}^m = (\lr A_{i,j}^{m+\half} + \lt A_{i,j}^m)\frac{\Delta t}{2},
\eeq
\beq
 A_{i,j}^{m+ 1} - A_{i,j}^{m+ \half} = (\lr A_{i,j}^{m+\half} + \lt A_{i,j}^{m+1})\frac{\Delta t}{2}.
\eeq
Here $\lr$ and $\lt$ are the difference forms of the operators $L_r$ and $L_\theta$.  They
have the forms
\beq
\lr A_{i,j}^m = a(i,j) A_{i-1,j}^m + b(i,j) A_{i,j}^m + c(i,j) A_{i+1,j}^m,
\eeq
\beq
\lt A_{i,j}^m = d(i,j) A_{i,j-1}^m + e(i,j) A_{i,j}^m + f(i,j) A_{i,j+1}^m.
\eeq
To obtain $a(i,j)$, $b(i,j)$, $c(i,j)$, $d(i,j)$, $e(i,j)$ and $f(i,j)$, one has to use appropriate
difference schemes for the operators $L_r$ and $L_\theta$.  We use the {\it Lax-Wendroff}
scheme for the advective part, whereas the diffusive terms
involving $\pa^2 A/\pa r^2$ and $\pa^2 A/\pa \theta^2$ are handled by a simple Centred Space
scheme (see, for example, Press {\it et al.\ } 1988). Since these diffusive terms are treated implicitly
in one half-step and explicitly in the next half-step, what we get is essentially a {\it Crank-Nicholson}
scheme for these diffusive terms.
The terms involving $\pa A/\pa r$ and $\pa A/\pa \theta$ in the diffusive part
are handled by an upwind scheme.
The expressions of $a(i,j)$, $b(i,j)$, $c(i,j)$,
$d(i,j)$, $e(i,j)$ and $f(i,j)$ are
given below:
\ber
a(i,j) &=& \frac{\eta}{(\Delta r )^2} + \frac{U_r (i,j)} {\Delta r} (r - \frac{\Delta r}{2}) \nonumber \\
&& \times \left[\frac{1}{2} + \frac{\Delta t}{4 \Delta r} U_r(i-\half,j) (r - \Delta r) \right], \\
b(i,j) &=&  -\frac{2 \eta}{(\Delta r )^2}  - \frac{2 \eta}{r \Delta r} - \frac{\eta}{2 r^2 \sin^2 \theta} 
            - \frac{U_r (i,j)} { \Delta r} \nonumber \\
&& \times \left[\frac{\Delta r}{2}  
+ \frac{\Delta t}{4 \Delta r} r \left\{U_r(i+\half,j)(r + \frac{\Delta r}{2}) \right. \right. \nonumber \\
&& \left. \left. + U_r(i-\half,j)(r - \frac{\Delta r}{2}) \right\} \right], \\
c(i,j) &=&  \frac{\eta}{(\Delta r )^2}  + \frac{2 \eta}{r \Delta r} - \frac{U_r (i,j)} { \Delta r} (r + \frac{\Delta r}{2}) \nonumber \\
&& \times \left[ \frac{1}{2} - \frac{\Delta t}{4 \Delta r} U_r(i+\half,j) (r + \Delta r) \right], \\
d(i,j) &=&  \frac{\eta}{(r \Delta \theta )^2} 
+ \frac{U_{\theta} (i,j)} {\Delta \theta} \sin(\theta - \frac{\Delta \theta}{2}) \nonumber \\
&& \times \left[ \frac{1}{2} + \frac{\Delta t}{4 \Delta \theta} U_{\theta}(i,j-\half) \sin(\theta - \Delta \theta) \right], \\
e(i,j) &=&  -\frac{2 \eta}{(r \Delta \theta )^2} - \frac{ \eta \cot \theta}{r^2 \Delta \theta} 
- \frac{\eta}{2 r^2 \sin^2 \theta} \nonumber \\
&& - \frac{U_{\theta} (i,j)}{ \Delta \theta} \nonumber \\
&& \times \left[ \sin(\theta + \frac{\Delta \theta}{2}) 
\left\{\half +  \frac{\Delta t}{4 \Delta \theta} U_{\theta}(i,j+\half) \sin \theta \right\}  \right. \nonumber \\
&&\left.  -\sin(\theta - \frac{\Delta \theta}{2}) 
\left\{\half -  \frac{\Delta t}{4 \Delta \theta} U_{\theta}(i,j-\half) \sin \theta \right\} \right], \nonumber \\
\\
f(i,j) &=&  \frac{\eta}{(r \Delta \theta )^2} + \frac{ \eta \cot \theta}{r^2 \Delta \theta}
- \frac{U_{\theta} (i,j)} {\Delta \theta} \sin(\theta + \frac{\Delta \theta}{2}) \nonumber \\
&& \times \left[ \frac{1}{2} - \frac{\Delta t}{4 \Delta \theta} U_{\theta}(i,j+\half) \sin(\theta + \Delta \theta) \right],
\eer
where $U_r (i,j)$ and $U_{\theta} (i,j)$ respectively give the
values of $v_r/r$ and $v_{\theta}/(r \sin \theta)$ at the
grid point $i,j$.

The equations are solved on a $64 \times 64$ grid.  The boundary conditions are already discussed
in the text.  It was mentioned that the bottom boundary condition is to allow the magnetic field to be
carried freely through the bottom boundary, where the velocity is radially inward.  We advance
$A_{1,j}^m$ at the bottom by the simple upwind differencing scheme:
\beq
\frac{A_{1,j}^{m+1} - A_{1,j}^m}{\Delta t} = - v_r(1+ \half,j) \frac{A_{2,j}^m - A_{1,j}^m}{\Delta r}.
\eeq


\beb
\bi Ames W.~F., 1977, {\it Numerical Methods for Partial Differential Equations}. Academic Press
\bi Bhattacharya D., 1995, {\em X-Ray Binaries}, ed. Lewin W.~H.~G., van Paradijs J., 
    van den Heuvel E.~P.~J., Cambridge University Press, pp. 233-251
\bi Bhattacharya D., 1996, {\em Pulsars : Problems and Progress}, ed. 
    Johnston S., Walker M.~A., Bailes M., ASP Conference Series Vol.105, pp. 547-556
\bi Bhattacharya, D., 1999a, {\em Pulsar Timing, General Relativity and the Internal Structure of Neutron Stars}, ed.
    Arzoumanian Z., Van der Hooft F., van den Heuvel E.~P.~J., Koninklijke Nederlandse Akademie van Wetenschappen, Amsterdam,
    p. 235
\bi Bhattacharya D., 1999b, {\em The Neutron Star - Black Hole Connection}, 
      ed. Kouveliotou C., van Paradijs J., Ventura J., Kluwer, {\rm in print}
\bi Bhattacharya D., Datta B., 1996, MNRAS, 282, 1059
\bi Bisnovatyi-Kogan G.~S., Komberg B.~V., 1974, SvA, 18, 217
\bi Blandford R.~D., DeCampli W.~M., K\"onigl A., 1979, BAAS, 11, 703
\bi Brown E.~F., Bildsten L., 1998, ApJ, 496, 915
\bi Choudhuri A.~R., 1989, SoPh, 123, 217
\bi Choudhuri A.~R., 1998, {\it The Physics of Fluids and Plasmas: An Introduction
    for Astrophysicists}. Cambridge University Press
\bi Choudhuri A.~R., Dikpati M., 1999, SoPh, 184, 61
\bi Choudhuri A.~R., Gilman P.~A., 1987, ApJ, 316, 788
\bi Choudhuri A.~R., Sch\"ussler M., Dikpati M., 1995, A\&A, 303, L29
\bi Cumming A., Zweibel E.~G., Bildsten L., 2001, astro-ph/0102178
\bi Dikpati M., Choudhuri A.~R., 1994, A\&A, 291, 975
\bi Dikpati M., Choudhuri A.~R., 1995, SoPh, 161, 9
\bi Dikpati M., Charbonneau P., 1999, ApJ, 518, 508
\bi Durney B., 1995, SoPh, 160, 213
\bi Durney B., 1997, ApJ, 486, 1065
\bi Emiliani C., 1992, {\it Planet Earth}.  Cambridge University Press
\bi Geppert U., Urpin V.~A., 1994, MNRAS, 271, 490
\bi Geppert U., Page D., Zannias T., 1999, A\&A, 345, 847
\bi Hameury J.~M., Bonazzola S., Heyvaerts J., Lasota J.~P., 1983, A\&A, 128, 369
\bi Jahan Miri M., Bhattacharya D., 1994, MNRAS, 269, 455 
\bi Konar S., 1997, PhD Thesis, IISc, Bangalore
\bi Konar S., Bhattacharya D., 1997, MNRAS, 284, 311
\bi Konar S., Bhattacharya D., 1999, {\em The Neutron Star - Black Hole Connection}, 
      ed. Kouveliotou C., van Paradijs J., Ventura J., Kluwer, {\rm in print}, astro-ph/9911239
\bi Konenkov D., Geppert U., 2000, MNRAS, 313, 66
\bi Konenkov D., Geppert U., 2001a, MNRAS, 325, 426
\bi Konenkov D., Geppert U., 2001b, A\&A, 372, 583
\bi Moreno-Insertis F., 1983, A\&A 122, 241
\bi Nandy D., Choudhuri A.~R., 2001, ApJ, 551, 574
\bi Parker E.~N., 1979, {\it Cosmical Magnetic Fields}. Clarendon Press
\bi Press W.~H., Flannery B.~P., Teukolsky S.~A., Vetterling W.~T., 1988, {\it Numerical Recipes}. Cambridge
	University Press
\bi Romani R. ~W., 1990, Nat, 347, 741
\bi Romani R. ~W., 1995, {\em Isolated Pulsars}, ed. van Riper K., Epstein R., Ho C., Cambridge University Press, pp. 75-82
\bi Spruit H.~C., 1983, {\em Solar and Stellar Magnetic Fields : Origins and Coronal Effects},
    IAU Symposium No. 102, ed Stenflo J.~O., D. Reidel Publishing Company
\bi Taam R.~E., van den Heuvel E.~P.~J., 1986, ApJ, 305, 235
\bi Urpin V.~A., Geppert U., 1995, MNRAS, 275, 1117
\bi Urpin V.~A., Geppert U., 1996, MNRAS, 278, 471 
\bi van den Heuvel E.~P.~J., 1995, JA\&A, 16, 255
\bi Verbunt F., van den Heuvel E.~P.~J., 1995, {\em X-Ray Binaries}, ed. Lewin W.~H.~G., van Paradijs J., van den Heuvel E.~P.~J., 
    Cambridge University Press, pp. 457-494
\bi Woosley S.~E., Wallace R.~K., 1982, ApJ, 258, 716
\eeb

\end{document}